\documentclass[apjl]{emulateapj}
\usepackage{tipa}
\usepackage[figuresright]{rotating}
\usepackage{subfigure}
\usepackage{epic,eepic}
\usepackage{graphicx}
\usepackage{longtable}
\usepackage{threeparttable}
\usepackage{appendix}
\usepackage{amsmath}
\usepackage{multirow}
\usepackage{color}
\shorttitle{The Flare Catalog and the Flare Activity in the {\it Kepler} Mission }

\shortauthors{Yang \& Liu}

\begin{document}

\title{The flare catalog and the flare activity in the {\it Kepler} mission}

\author{Huiqin Yang\altaffilmark{1}, Jifeng Liu\altaffilmark{1,2,3}}

\altaffiltext{1}{Key Laboratory of Optical Astronomy, National
Astronomical Observatories, Chinese Academy of Sciences, Beijing
100101, China; yhq@nao.cas.cn}

\altaffiltext{2}{School of Astronomy and Space Sciences, University of Chinese Academy of Sciences, Beijing 100049, China}

\altaffiltext{3}{WHU-NAOC Joint Center for Astronomy, Wuhan University, Wuhan, Hubei 430072, China}
\begin{abstract}
We present a flare catalog of the {\it Kepler} mission using the long-cadence data of Data Release 25. This catalog comprises 3420 flare stars and 162,262 flare events. A comparison shows that the flare catalogs of previous studies are seriously polluted by various false positive signals and artifacts. The incidence of flare stars rises with decreasing temperature, which accords with the theoretical analysis. The flare frequency distributions (FFDs) from F-type stars to M-type stars obey a power-law relation with $\alpha \sim 2$, indicating that they have the same mechanism on generating flares. The remarkable incidence and the deviation of FFDs on A-type flare stars imply that they generate flares in a different way. The activity--rotation relation is consistent with previous studies at low temperature band, whereas it becomes dispersive with increasing temperature. Combined with the Gyrochronology, we find that the mixing of stars of two different dynamos gives rise to the dispersion. We thereby propose a scenario on understanding the activity--rotation relation across the H-R diagram. Based on the scenario and the correspondence of dynamo with regard to activity and rotation, we suggest a new expression on the activity--rotation relation, in which the segmentation is on the basis of the dynamo rather than the rotation period. The rotation distribution of flare stars shows that about 70\% of flare stars rotate faster than 10 days and the rate approaches 95\% at 30 days. Based on the incidence and the rotation distribution of flare stars, we estimate that a superflare with energy $\sim 10^{34}$ erg occurs on the Sun at least once in 5500 years.

\end{abstract}

\keywords{Stars:flare --- stars: activity --- stars: rotation --- stars: evolution --- stars: statistics}

\section{INTRODUCTION}
A flare, accompanied by particle beams, chromospheric evaporation and stellar mass ejections, is the most prominent manifestation of stellar activity. It can be observed in a wide range wavelengths and could cause huge change both on photometry and spectrum \citep[e.g.,][]{Benz2010}. On our Sun, a typical flare lasts several minutes with energy ranging from $10^{29}-10^{32}$ erg, which can be well supported by the classical model of the solar magnetic reconnection.

Stellar flares are also frequently detected and extensively studied. Given most of our concepts on stellar activity and stellar structure are from the Sun, it is a natural thought that stellar flares have the same mechanism. Their profiles and flare frequency distributions (FFDs) are similar to solar flares \citep{Maehara2012,Shibayama2013,Yang2017}. However, there are some differences between them. For example, energies of stellar flares may be much larger than that of solar flares \citep{Walkowicz2011,Maehara2012}, while durations of stellar flares are much shorter than expected values from the solar observations \citep{Name2017}. White light flares are easier to observe, whose spectral energy distributions(SEDs) are often described by a blackbody spectrum. For very active stars (e.g., dMe stars), the temperature of the blackbody is about 9000 K \citep[e.g.,][]{Kowa2013,Hawley1992}, while for the Sun, the accurate value of the temperature is in dispute. \citet{Kretzschmar2011} reported 9000 K that is similar to stellar flares, whereas \citet{Kleint2016} argued that it is about 6000 K by analysing the influence of the pre-flare continuum. Moreover, the excess energies in the near-UV band indicate that there should be an extra component to fit the SEDs \citep{Kleint2016,Kowa2019}.

One straight connection to the flare study is stellar activity, whereas the Sun is a relatively quiet star. The study of active stars thus plays an important role in the realm. The classical indicator for stellar activity is spectral emission lines such as  Ca~{\sc ii} and H$\alpha$ that correlate each other well \citep{Fang2018}. The components from the chromosphere are believed to depend on the magnetic activity, which leave many clues on stellar structure and evolution. For example, the long term variation of Ca HK lines reflect the stellar cycle \citep{Saar1999,Metcalfe2016}. The intensities of those lines are also applied to predict stellar age \citep[e.g.,][]{Soder1991,Mamajek2008}.

However, it is difficult to grasp a comprehensive physical picture across H-R diagram through chromospheric emission lines, because they may be hard to measure and may not always be present in different stellar types and different stellar activities. Moreover, it is complicated to unify various indicators of emission lines such as S index, $R_{\rm HK}$ and $R'_{\rm HK}$ \citep[e.g.,][]{Noyes1984,Soder1991,Egeland2017}, which are calibrated by the instruments of Mount Wilson Observatory.

The X-ray luminosity is from the corona of a star. It is used to investigate stellar activity \citep[e.g.,][]{Pizzolato2003,Wright2011,He2019}. However, it is confronted with the matter of unity and coherence as well, because mechanisms on generating coronal X-ray luminosity are different \citep{Lucy1980,Vaiana1981}, i.e., the X-ray luminosity of the hottest and most massive stars are from either shocks in their winds or collisions between the wind and circumstellar material \citep{Parkin2009}, while that of the late type star is caused by the magnetic corona. Furthermore, some issues such as the coronal heating are still poorly understood, in spite of that they depend on the magnetic activity.

The flare, as another indicator, is closely related to the magnetic activity \citep{Kowa2009,He2018}. It is easy to measure and can be widely observed in different spectral types. It is also better at disclosing the activity variation on a short time scale, given that emission lines vary on a time scale of years. However, the constraint of the flare study is equally obvious. Since there is no way of predicting a flare in advance, a continuous and accurate observation is required to change the situation of serendipitous detections.

The launch of the Kepler spacecraft \citep{Borucki2010} has opened a new era in the study of flares, which allows for the detection of hundreds and thousands of flares and with improved accuracy. The progress is not only on number. The {\it Kepler} mission also detects new flares that leads to new scientific questions. For example, Superflares found in solar-like stars raise a question that whether they could occur on our Sun \citep{Maehara2012}, which is interesting to the public \citep{Schri2012}. In past years, independently flaring events in X-ray band have been reported on early type stars \citep[e.g.,][]{Smith1993,Groote2004}, which do not have efficiently deep convective zones. They were usually interpreted by other mechanisms in order to avoid a theoretical conflict \citep{Townsend2005,Capelli2011}. However, it is incredible that all of the frequent flares in early type stars found by {\it Kepler} are due to the accretion or the shock of stellar wind \citep{Balona2013,Pedersen2017}. Quasi-periodic pulsations of flares found by {\it Kepler} could be explained by the MHD oscillation on the Sun \citep{Pugh2016}, which implies that stellar flares have a correspondence with solar flares.

Thanks to the {\it Kepler} mission, a systematic study of stellar flares became possible. It enables us to conduct a statistical study on flares, which could reveal the flare activity and the flare occurrence across H-R diagram in a uniform way. It is important for us to grasp a whole picture and verify the importance of various factors for stellar activity.

On a wider scope of the study of stellar activity, stellar activity associates many aspects of stellar physics. when a star evolves, its rotation gradually becomes slower because it loses angular momentum through magnetized stellar wind \citep{Gallet2013}. Stellar wind depends on stellar dynamo and stellar structure, which also determine stellar activity. On the other hand, The decline of the rotation reacts the stellar dynamo and the magnetic field configuration. It hence restrains stellar activity. The coupling connections are gradually summarized and further induced as relations or indicators. For example, the activity-rotation relation confirms the theoretical analysis, \citep[e.g.,][]{Skumanich1972,Noyes1984,Pizzolato2003,Wright2011,Yang2017}, which can shed light on stellar dynamo, structure and age. Although there are still some issues to be addressed \citep[e.g.,][]{Douglas2014,Wright2016}, one can expect to see the location of different stars with different ages and structures in this relation. The age--rotation relation (also be dubbed as Gyrochronology) develops a measurement of age for the main sequence star through the rotation period, and classifies tracks of stellar evolution with different dynamos and structures \citep{Barnes2003, Barnes2007, Barnes2010,Meibom2015}.

The study of stellar activity involves many realms of stellar physics and consequently has different forms. However, the core of those issues is that they are closely related to stellar structure and evolution. It is thus inevitable and important to aim flare study at those directions. The large sample of the {\it Kepler} mission provides us an opportunity to gauge those issues from the view of flares.

In Section 2, we introduce the {\it Kepler} mission and data. We then present methods on the flare detection, the energy estimation and the calculation of the flare activity. In Section 3, the flare catalog is listed, accompanied by comparison and analysis on the catalog. In Section 4, we investigate and discuss issues on the activity, the rotation period, the age, the dynamo and their relations. We give our conclusions in Section 5.
\section{Data and Method}

\subsection{The Kepler Data}
With the main goal of exploring exoplanets, {\it Kepler} spacecraft was launched by NASA in 2009. It can monitor over 150,000 targets simultaneously by a large field of view of 105 $deg^2$. High precision and continuity are the most significant character of the {\it Kepler} mission. Its precision for bright targets (V = 9-10) approaches 10 ppm, and 100 ppm even for faint targets (V = 13-14). Its capacity of sampling interval is up to 1 minute for short cadence (SC) data and 30 minutes for long cadence (LC) data \citep{Van Cleve2009}.

Another advantage of {\it Kepler} is that physical parameters of most targets such as the effective temperature are obtained through multi-color photometry in advance and listed in Kepler Input Catalog \citep[KIC;][]{Brown2011}, which provide great convenience to the research of the {\it Kepler} data. Although there are some disputes between KIC and stellar theory for the early type and M-type stars \citep[e.g.,][]{Dressing2013, Batalha2013}, the continuous versions have revised the inconsistence to a certain extent by various methods \citep{Huber2014,Mathur2017}.

The lifetime of the {\it Kepler} mission is divided into two stages. The first stage (K1) is from 2009 to 2013, when its telescope continuously pointed at a single star field in Cygnus-Lyra region. Limited by the efficiency of transportation, K1 finally provided data for about 200,000 stars, whose lightcurves are up to 17 quarters.

Due to the out of control of the reaction wheels, {\it Kepler} began the second mission (K2) to observe around the ecliptic plane in 2014. K2 is a new mode of observation because its scientific agendas are community-driven and the observational time for each campaign is short. The quality of the data of K2 is not as good as K1 \citep{Ilin2019}. Its targets often slowly shift on CCDs and instantaneously return to original positions, leading to spikes on lightcurves \citep{Vancleve2016}. It is more difficult to use them to carry out a study on the short-time variation, especially on the statistic study of flares. Our work is thus based on K1.

\citet{Yang2018} found 940 flares in SC data. However, SC data only have about 5,000 targets and their average observational time is about two months. In this work, we detect flares in LC data, which include over 200,000 targets. Our detection covers the whole data set (Q1-Q17; 48 months; Data Release 25). The uncorrected simple aperture photometry (SAP) data are utilized in our study, as is done in \citet{Balona2015}, \citet{Davenport2016} and \citet{Yang2017}.

\subsection{Detection Method}
\begin{figure*}[!htb]
\begin{center}
\subfigure{\includegraphics[width=1\textwidth]{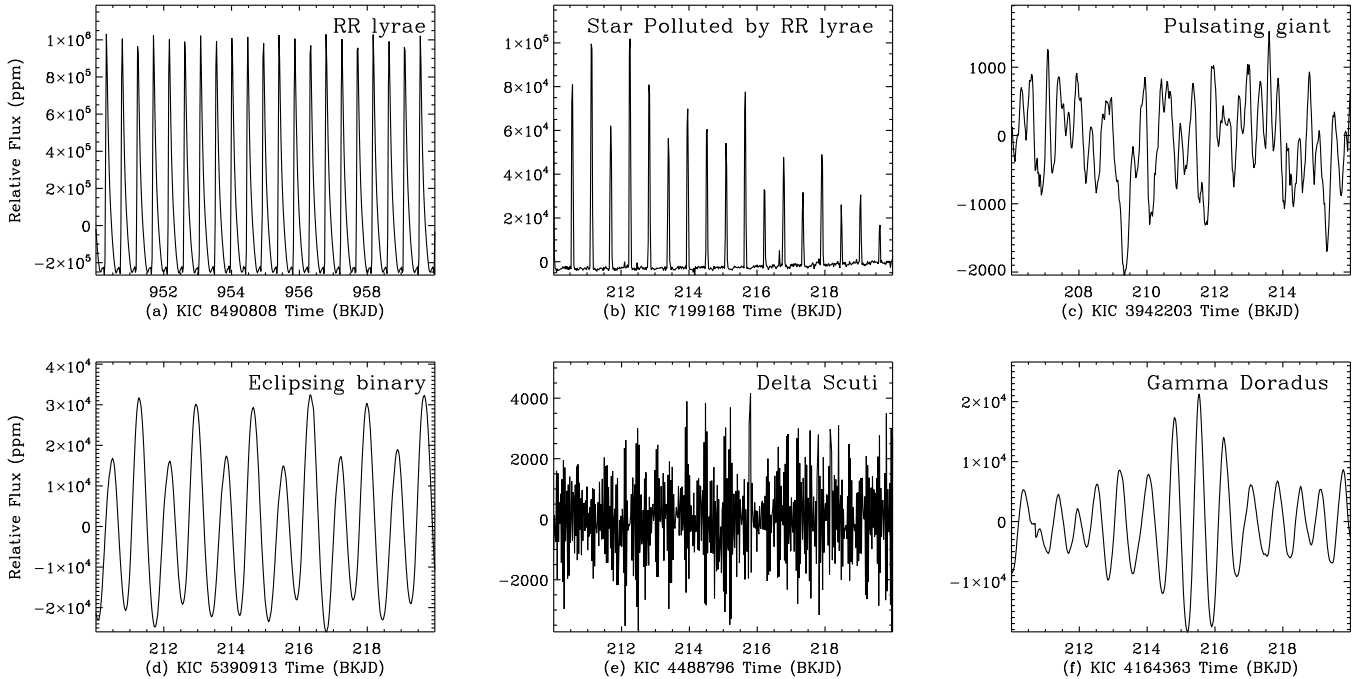}} \hspace{1
mm}

\caption{Examples of various pulsating stars or pollution. They are RR lyrae, star polluted by RR lyrae, pulsating giant, eclipsing binary, $\delta$ Scuti, and $\gamma$ Doradus from panel (a) to panel (f) respectively. Their number is numerous and they can cause either spikes or dramatically changing baselines on lightcurves. Those features can result in false positive signals. } \label{figctn}
\end{center}
\end{figure*}
For most studies on the flare detection \citep{Walkowicz2011,Hawley2014,Gao2016,Davenport2016,Silv2016,Door2017,Yang2017,Vida2018}, the detection method can be mainly summarized into three steps:

(1)The first is to fit the quiescent flux or the baseline. It includes a prewhitening procedure to remove noises and potential artifacts in raw data and an iteratively smooth filter to rule out outliers, in which a low-order, slowly changing function is often used.

(2)After detrending the baseline, different criteria could be applied to outliers to identify whether they are flares. In spite of the difference, all of them concern the amplitude, the profile and the duration. For example, a common approach is to set a threshold on the amplitude that should be larger than 3$\sigma$. The profile should have an impulsive rise (relatively short) and an exponential decay (relatively long). The number of continuous points that represent the duration should be set to separate flares from pollution such as cosmic rays.

(3)The last is to validate the result and check pollution. It often refers to mathematic tests or simulations, which are expected to give a quantitative confine on selection criteria and errors. However these methods seem to merely compress the whole sample and do not distinguish flares and artifacts, because they cannot make up the potential deficiency of the detection functions. For example, \citet{Davenport2016} used the flare equivalent duration (ED) to test the recovery efficiency by injecting artificial flares, then he imposed a threshold of the ED and the flare number on the selection criteria. We check his catalog and find that it is hard to prevent pollution. We will discuss this comparison in detail in Section 3.

Based on the previous studies, we present our detection method step by step for clarity.

(1) Inner discontinuities (longer than 6 hours) of each quarter are divided into blocks and each block is fitted independently. In a block, a third-order polynomial is applied to remove instrumental effects. The instrumental effects are long-term trends. In general, they cannot account for starspot variations, but can influence the accuracy of the Lomb-Scargle periodogram. Then the Lomb-Scargle periodogram is calculated for each block to ascertain the most significant frequency. A smoothing filter that based on the spline algorithm \citep{Press2007} is used to fit the baseline. The filter width is one fourth of the most significant period. Its lower limit is 6 hours and upper limit is 24 hours \citep{Yang2017}. In the fitting process, an iterative $\sigma$-clipping approach is necessary to remove all the outliers. The starspot variations are removed by fitting baselines.

(2) The criteria to identify flare candidates are as follows:

(i) \citet{Yang2018} found that most of flare candidates with one or two points in LC data were artifacts. Therefore, after detrending baselines, at least three continuous points higher than 3$\sigma$ are sorted out as a flare candidate and all the candidates must have no breakpoints within 3 hours. Its end is at the last continuous point that is higher than 1$\sigma$.

(ii) Some previous works defined the profile of a flare by a function \citep{Pugh2016,Door2017}. However, \citet{Yang2018} found that LC data were hard to capture the profile of flares by comparing them with their counterparts of SC data. It hence is suitable to set a relatively loose criterion on the flare profile, i.e., the decay phase should be longer than the rise phase.

(iii) If a target has multiple quarters, its flares should occur at least in two quarters.

(iv) If a target has a unique candidate, its energy should be larger than $10^{34}$ erg and its duration should be at least 2 hours.

(3) Strictly speaking, the robovetting procedure is
mainly to filter out artifacts rather than to identify true flares. In fact, a considerable part of candidates are false positive signals or are subjected to pollution, although they meet all of the above criteria. Figure~\ref{figctn} shows representative false positive signals.

The false positive signals are mainly attributed to two folds. One is from the pulsating star, whose pulsating profile is similar to the profile of flares. The pulsating stars consist of $\delta$ Scuti, $\gamma$ Doradus, pulsating giants, RR lyrae and Cepheid in the {\it Kepler} field and their number is enormous. They could cause either sharp peaks or dramatically changing baselines. For example, as shown in the panel (e) of Figure~\ref{figctn}, a $\delta$ Scuti star could give rise to spikes in LC data by high frequency oscillations of the p mode. Panel (a) of Figure~\ref{figctn} shows a radial pulsating star RR lyrae, which has evolved off the main sequence. Its pulsating profile has a quick rise and a slow decay with a high amplitude, which is very analogous to that of flares. We refer to \citet{Aerts2010} for mechanisms of pulsating stars and \citet{Balona2011a,Balona2011b} for their morphology in the {\it Kepler} data.

It is hard to separate them from flares thoroughly by automatical routines. For example, we checked the catalog of \citet{Davenport2016} and find it was seriously polluted by various pulsation stars. \citet{Door2017} tried to use a method that is similar to inverse fast fourier transform (IFFT) to remove pulsating stars, but we still found many pulsating stars in their sample. In our sample, many pulsating stars have been excluded in the the robovetting procedure and more than 3,000 pulsating stars are excluded out of about 8300 candidate stars by visual inspection.

Another reason for the artifacts is the instrumental error. For example,\citet{Yang2018} found that in some fixed time segments, analogous artifacts appeared repeatedly, which could be caused by charge transfer efficiency \citep{Coughlin2016}. Candidates in those segments are excluded.

The validation of the results mainly depends on visual inspection. We have developed an online platform: Kepler Data Integration Platform (KDIP)\footnote{http://kepler.bao.ac.cn}, which integrates query, view, calculation and categorization on the whole {\it Kepler} data set. The platform is based on the network language JavaScript, PHP, and the database language MySQL. One can instantaneously perform the detection procedure for each target and dynamically zooms in or out to check flare candidates on a lightcurve. It enhances the efficiency greatly for us to complete the tremendous work. It also provides remarkable help on the contamination check, which we will discuss in Section 2.3.

In the inspection, a small part of lightcurves are refitted by finely tuning the filter width, especially for targets with ultra-rapid rotation \citep{Yang2017}. Some marginal candidates are deemed as flares with arbitrary choices that are based on our experiences. The overall quality of a lightcurve is taken into account as well. Figure~\ref{figctn} shows spurious signals by pulsating stars and Appendix A illustrates some typical deceptive cases in the flare detection. Over 10,000 targets are vetted by eye.

\subsection{Contamination Check}

The typical photometric aperture of the {\it Kepler} mission has a radius of 4--7 pixels \citep{Bryson2010}, It is thus common for a given target to be contaminated by nearby objects, given that some sources are very close to each other on CCDs. About 10\% of flare candidates are probably polluted due to this reason \citep{Shibayama2013, Gao2016}. In order to avoid these false events, we checked for possible contamination as follows:

(i) 152 flaring stars have adjacent field stars located within 12\arcsec . They are possibly polluted by neighboring stars. However, we flag them rather than exclude them.

(ii) The {\it Kepler} eclipsing binary catalog
(KEB)\footnote{http://keplerebs.villanova.edu/} includes more than 2800 eclipsing binaries. 135 of them
are removed from our sample. A flare study on the binary sample is referred to \citet{Gao2016}.

(iii) The check of centroid offset are done. About
2000 flaring events are removed in this step. We refer to \citet{Yang2017} for a detailed description of this step.

\subsection{Flare Activity}
The energy of each flare is calculated in our catalog. The approach of the energy estimation is according to \citet{Shibayama2013}
and \citet{Yang2017}. Its main principles are as follows: (1)
A white light flare is assumed to radiate as a blackbody with effective temperature of 9000K per unit area. (2) The flare
area is estimated by the flare amplitude, the stellar radius, the stellar
effective temperature, and the {\it Kepler} response function.

The estimation of the flare energy depends on the flare temperature, whose accurate value is in dispute \citep{Kretzschmar2011,Kleint2016}. However, we should note that regardless of whether the flare temperature is 9000 K or 6000 K, the difference of the flare energy is $E_{\rm 9K} / E_{\rm 6K}-1$ = 22.7\%. This is smaller than the error of the flare energy, which is about 60\% \citep{Shibayama2013}. Furthermore, we stress that the the energy of the white light flare is obtained in the optical band, since the SEDs in this band can be well fitted by a blackbody model. However, the blackbody model does not explain the NUV, which exhibits a Balmer jump. In the NUV, it appears excess energies that need extra components that could be from different layers to explain \citep{Kowa2019}.

We then normalize flaring energies to flare activities for each star, i.e., $\sum E_{\rm flare}/\int L_{\rm bol}dt = L_{\rm flare}/L_{\rm bol}$ \citep{Yang2017}. As other indicators (e.g., $L_{\rm x}/L_{\rm bol}$), the normalization can represent active level in a unified way by removing the influence of the stellar luminosity. Note that $L_{\rm flare}$ in the ratio is the average flaring luminosity. This quantity in the U band is found to be well correlated with the stellar luminosity \citep{Lacy1976,Osten2012}.

\citet{Yang2018} searched all the flares in SC data and compared them with their counterparts in LC data. They found that LC flares could capture the basic properties of flares, while LC energies are underestimated by 25\%. We have compensated for 25\% of LC energies to calculate the flare activity.

\section{The Flare Catalog}
In total, we have identified 3420 flare stars out of over 200,000 targets in the {\it Kepler} data. The number of flare events is 162,262. Table 1 lists all of the flare stars including the flare number, the flare activity and the rotation period, of which 152 targets that have neighboring stars are labeled. Note that the periods of 2492 stars are according to \citet{Mc2014} and \citet{Reinhold2013}, and the rest are calculated by Lomb-Scargle algorithm. 291 stars do not show apparent modulations of lightcurves caused by starspots. Table 2 presents all of the flaring events with the begin time, the end time and the energy.

\begin{table}
\renewcommand{\arraystretch}{1.5}
\begin{center}
\caption[]{ parameters of flaring stars } \vspace{0.1mm}
\label{table1}\small \tabcolsep=3.pt
\begin{threeparttable}
\begin{tabular}{cccccccccc}
\tableline \tableline KIC ID & $N_{\rm flare}$ & $T_{\rm eff}$ & log$g$ & $L_{\rm fl}/L_{\rm bol}$
 &$P_{\rm rot}$ & NS\\
 & & (K) & ($cm \cdot s^{-2})$ & & (Day) & \\
\hline

8935655 &491& 3694&4.69&$5.22 \times 10^{-7}$&2.00\tnote{a}&No\\
9201463 &489& 3319&5.14&$1.83 \times 10^{-6}$&5.55\tnote{c}&No\\
8093473 &438& 3360&4.96&$5.10 \times 10^{-7}$&6.04\tnote{a}&No\\
4248763 &410& 4353&4.57&$4.92 \times 10^{-7}$&1.79\tnote{a}&No\\
10332732 &402& 3310&5.01&$5.10 \times 10^{-7}$&3.49\tnote{a}&No\\
7131515 &401& 3838&4.65&$6.58 \times 10^{-7}$&3.86\tnote{a}&No\\
7117293 &395& 3377&4.95&$4.22 \times 10^{-7}$&2.73\tnote{a}&No\\
10857583 &394& 3577&4.94&$4.88 \times 10^{-7}$&10.97\tnote{a}&No\\
9540467 &332& 3919&4.68&$1.92 \times 10^{-7}$&8.56\tnote{b}&No\\
12646841 &309& 3344&4.98&$8.38 \times 10^{-7}$&3.25\tnote{b}&Yes\\
\tableline
\end{tabular}

    \begin{tablenotes}
       \item[] Notes: Parameters of the 3420 flaring stars. The table lists the KIC ID, the flare number, the effective temperature, the surface gravity (log$g$), the flare activity and the rotation period. The flag `NS' indicates whether the star has neighboring stars. The effective temperature and the surface gravity are given by KIC. This table is available in its entirety online.
 \vspace{2mm}
        \item[a] Given by McQuillan et al. (2014).
        \item[b] Given by Reinhold et al. (2013).
        \item[c] Given by this work.
      \end{tablenotes}
\end{threeparttable}
\end{center}
\end{table}

\begin{table}
\renewcommand{\arraystretch}{1.5}
\begin{center}
\caption[]{ parameters of flares } \vspace{1mm}
\label{table1}\small \tabcolsep=5.pt
\begin{threeparttable}
\begin{tabular}{cccccccccc}
\tableline \tableline KIC ID & Quarter & Begin time & End time & Energy\\
 & &(Day)& (Day) &(erg)\\
\hline

8935655 &10& 907.7656&907.8269&33.684\\
8935655 &10& 908.6034&908.6647&32.7958\\
8935655 &10& 910.3199&910.3608&32.7446\\
8935655 &10& 911.0759&911.1985&34.2288\\
8935655 &10& 911.2803&911.4233&33.4663\\
8935655 &10& 912.7515&912.8333&33.1683\\
\tableline
\end{tabular}

    \begin{tablenotes}
       \item[] Notes:The table presents the details of the 162,262 flare events. It includes the KIC ID, the quarter, the begin time, the end time and the energy. The energy is in logarithm. This table is available in its entirety online.

      \end{tablenotes}
\end{threeparttable}
\end{center}
\end{table}
\begin{figure*}[!htb]
\begin{center}
\subfigure{\includegraphics[width=0.95\textwidth]{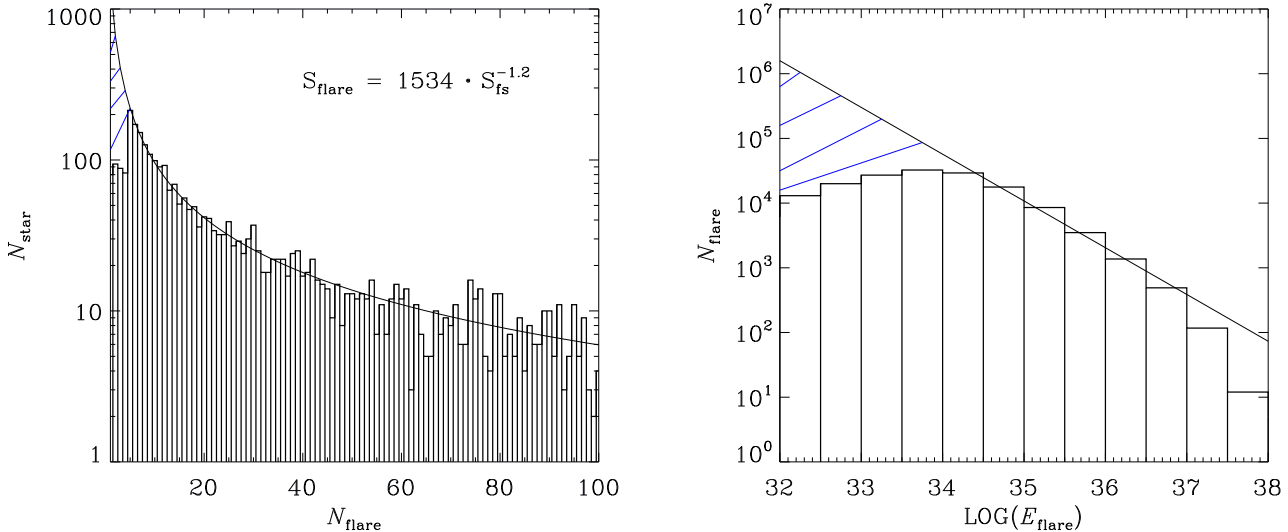}} \hspace{1mm}

\caption{The estimation of the lost rate on the flare stars and the flare energies. In the left panel, the $x$ axis is the given flare number for a star. The $y$ axis is the number of flare star in that given flare number. The statistical results in both panels are fitted by a power-law function. The blue oblique lines mark the lost region. The left panel indicates that about 2500 inactively flaring stars could be missed, consisting of about 5000 small flares. The right panel indicates that about 7\% flaring energies could be lost within the detection limit of the {\it Kepler} mission.} \label{figest}
\end{center}
\end{figure*}
\subsection{Comparison}

 It is interesting to investigate reliability and completeness of our catalog by comparing with previous studies. However, there are only a few flare studies oriented to the whole {\it Kepler} data set, and the lack of specific information on each flaring event increase the difficulty of the comparison.

\citet{Davenport2016} searched flares on LC data of Data Release 24 by fully automatic routines. He listed 4041 flaring stars without detailed information of each flaring event. The overlaps of the two catalogs are only 396 targets, which means 3645 targets of his catalog are not in our sample. In fact, they are various pulsating stars, fast rotating stars and stars with transit events. Those populations could cause either spikes or rapidly changing baselines leading to false positive signals. On the other hand, 3024 targets of our sample are not reported by him, because he merely listed stars with more than 100 flares.

\citet{Door2017} used LC data of quarter 15 to detect flares. In total, they have detected 16,850 flares on 6662 stars. 3028 flaring stars of our sample have LC data of quarter 15. 2223 targets are both in our subsample and their catalog leaving 4439 targets out of our sample. Most of them are polluted by the same reason mentioned above. However, a small portion of them may meet most of our criteria and look exceedingly like true flares, although we deemed them as artifacts in the robovetting procedure or the validation. Appendix B presents typical cases of a detailed comparison.

805 targets of our sample are not reported by \citet{Door2017} consisting of many very active stars. It may be because of the lack of quarter 15 of their data, given that the version of Data Release is not declared in their work.

It makes little sense to give a further quantitative comparison, considering different data range and version, but the comparison reminds us that our catalog may miss some small flares and inactively flaring stars, since in the LC data, there is no clear and objective way to identify those marginal cases and a conservative policy is adopted for them. We could estimate the lost rate of our sample through the distribution of the flare number and the flare energy.

Figure~\ref{figest} shows the estimation of the loss rate of flare stars and their energies. Assuming that the flare number distribution obeys a power-law relation, we could estimate the lost number of inactive stars. The lost regions are marked by oblique lines. As shown in the left panel, about 2500 inactively flaring stars could be missed, consisting of about 5000 small flares. The detection limit of the {\it Kepler} mission for the faintest star (i.e., M-type stars) is about $10^{32}$ erg \citep{Yang2017}, which is set as the lower limit of the lost energies. As shown in the right panel, the lost energies are small. Their proportion over the whole flare energies is about 7\%, indicating that within the detection limit, the lost energies have little influence on the flare activity.
 \subsection{Incidence of Flare Stars}
\begin{table*}
\renewcommand{\arraystretch}{1.5}
\begin{center}
\caption[]{Incidence of Flare Stars across H-R Diagram } \vspace{1mm}
\label{table1}\small \tabcolsep=8.pt
\begin{threeparttable}
\begin{tabular}{cccccccccc}
\tableline \tableline  Type & $T_{\rm eff}$ & $N_{\rm star}$& $N_{\rm fs}$ & Incidence& Ba2015 & Ba2015 &Van2017&\\
 &(K) && &&(LC) &(SC)&\\
\hline

A &$>$7500&3190&37&1.16\% $\pm$ 0.19\%&2.78\%&2.36\%&1.31\%\\
F &6000--7500&66522&459&0.69\% $\pm$ 0.03\%&0.94\%&2.54\%&3.20\%\\
G &5000--6000&93973&1372&1.46\% $\pm$ 0.04\%&3.29\%&4.91\%&2.90\%\\
K &4000--5000&29730&880&2.96\% $\pm$ 0.10\%&\multirow{2}{*}{12.75\%}&\multirow{2}{*}{10.16\%}&\multirow{2}{*}{5.28\%}\\
M &$<$4000& 6898&672&9.74\% $\pm$ 0.38\%&~&~&~\\
Giant &log$g <$ 3.5 &23030 &76&0.33\% $\pm$ 0.04\%&--&--&2.86\%\\
\tableline
\end{tabular}

    \begin{tablenotes}
       \item[] Notes:The incidence of flare stars grouped by spectral type. The results of previous studies are also listed. $N_{\rm star}$ is the total star number, and $N_{\rm fs}$ is the number of flare stars. Ba2015 represents \citet{Balona2015}. Van2017 represents \citet{Door2017}. The incidence in K- and M-type is merged together in their studies. Error bars of our incidence are estimated using a square root of the number of flare stars, which is the 1$\sigma$ confidence range, by assuming the probability of detected number of flare stars follows the Poisson distribution.

      \end{tablenotes}
\end{threeparttable}
\end{center}
\end{table*}
 \begin{figure*}[!htb]
\begin{center}
\subfigure{\includegraphics[width=0.95\textwidth]{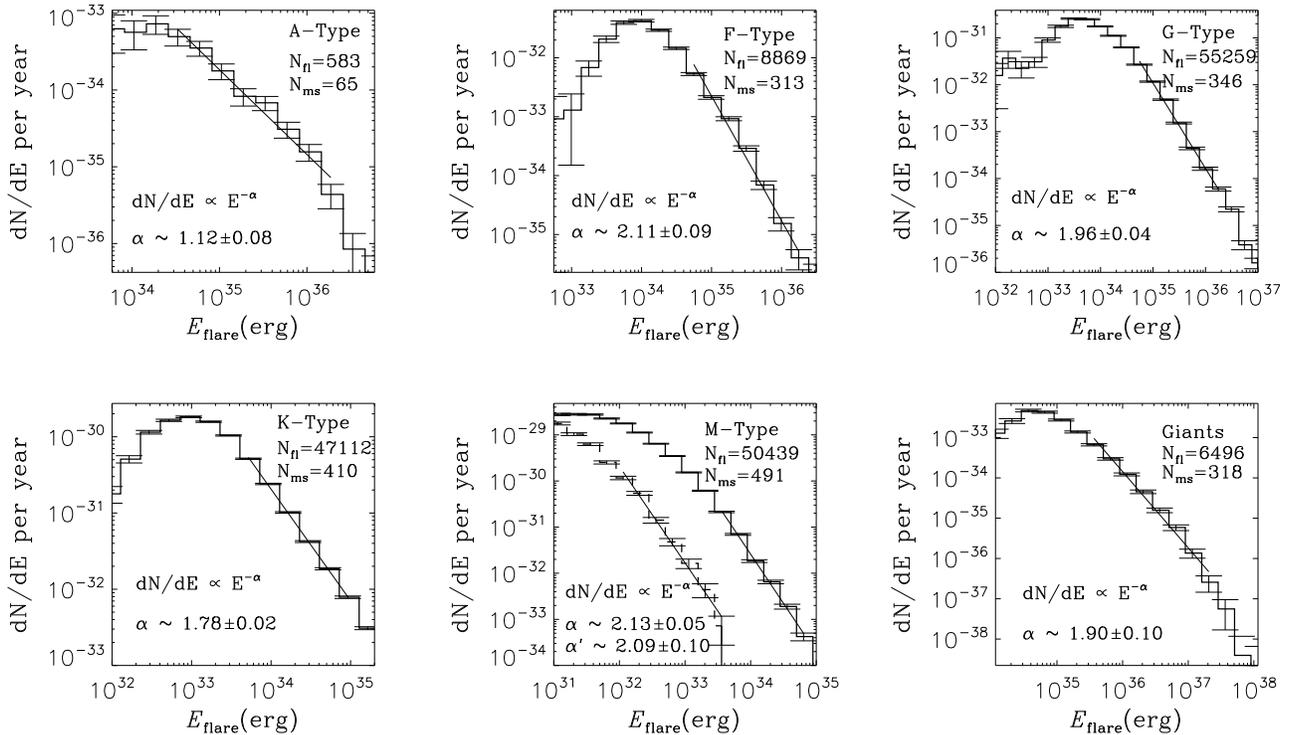}} \hspace{1mm}

\caption{The flare frequency distributions (FFDs) are grouped by the spectral type. An ordinary least-square (OLS) bisector is applied to fit the solid line \citep{Isobe1990}, whose slope is the $\alpha$ index. An energy range is set in each distribution. Energies within the range are used in the fitting. The range for A-,F-,G-type is from $10^{34.5}$ to $10^{36}$ erg, for K-type and M-type is from $10^{33.5}$ to $10^{35}$ erg and for giants is from $10^{35.5}$ to $10^{37}$ erg. $N_{\rm fl}$ and $N_{\rm ms}$ denote the flare number and the maximum number of flares per star in each spectral type respectively. The dashed line in the bottom middle panel shows the FFD of fully convective stars ($T_{\rm eff} < 3200 $K), whose index is 2.09. Error bars of each energy bin are estimated using a square root of the number of flares in each bin.} \label{figffd}
\end{center}
\end{figure*}
Table 3 shows the incidence of flare stars grouped by the spectral type. The results of \citet{Balona2015} and \citet{Door2017} are also listed for comparison. Error bars of our incidence are estimated using a square root of the number of flare stars, which is the 1$\sigma$ confidence range, by assuming the probability of detected number of flare stars follows the Poisson distribution. From the overall point of view, one can see that our results are much smaller than the previous works. With regard to \citet{Door2017}, the difference is mainly at F-type stars, G-type stars and giants, in which is filled with pulsating stars. Undoubtedly, the pulsating stars pollute the incidence seriously, and the pollution give rise to a spurious conclusion of the previous works that the incidence does not change apparently when the convection zone become deeper.

\citet{Balona2015} searched flares both on LC and SC data from quarter 1 to quarter 12 by visual inspection. Unfortunately, he merely lists SC flares without presenting any information of LC flares. His incidence is also much larger than our result. One possible reason of the difference is selection bias, because he sets a threshold that the {\it Kepler} magnitude $K_p < 12.5$. Flaring stars could be brighter than non-flaring stars on average for the same spectral type. Moreover, the sensitivity of SC data is much higher than that of LC data. It seems that the SC incidence should be greater than the LC incidence, whereas they are similar as shown in Table 3.

One should keep in mind that the incidence is not a constant but depends on the observational time, i.e., as shown in the left panel of Figure~\ref{figest}, the number of flare stars or the incidence of flare stars rises with increasingly observational time. Our results reflect the incidence of flare stars in 4 years. Also, we should note that, strictly speaking, what the result of Table 3 refers to is the incidence of superflare stars, because the flare detection depends on the detection limit and the quality of the data. It is inevitable that a lot of stars that only generate small flares are missed.

\citet{Yang2017} compared flaring stars and emission line stars in M dwarfs, finding that emission line stars are contained in flaring stars. All of the flaring stars with $P_{\rm rot} < 10$ days have emission lines, while others may not. They attributed the reason to the recovery time scale of the atmosphere from the emission level to the basal level after a star flared. Their relation will lead to that the flaring rate is higher than the emission rate in inactive stars and the two rates keep balance in active stars. This scenario is proved by M dwarfs where the activity has a dramatic rise with decreasing mass. The emission rate of early M dwarfs is about 2\% and over 20\% in middle and late M dwarfs\citep{West2004}. In this work, the flaring incidence is about 10\% in the early M dwarfs, and is the same as the emission rate of middle M dwarfs.

One interesting fact is that the incidence of A-type stars is higher than that of F-type stars both in our result and \citet{Balona2015}, and it monotonically increases after the F-type. From the A-type to the F-type, it includes the transition from a radiative envelope to a convective envelope in the exterior layer of a star. The traditional view on magnetic activity is that it is based on the convection zone, because the $\alpha-\Omega$ dynamo is powered by the tachocline that locates at the boundary between the radiative layer and the convective layer.  The interaction between magnetic flux tubes and the convection amplifies the magnetic field and makes magnetic flux tubes buoyant so that they can rise to the stellar surface \citep[e.g.,][]{Parker1975}. It is hence easy to understand the monotonous rise of the incidence after F-type stars because the depth of the convection zone increases. However, the discovery of flaring A-type stars and the remarkable incidence are confusing, unless we assume that they have a different mechanism of generating flares.

\citet{Balona2012,Balona2013,Balona2015} continuously reported more than 50 flaring early type stars. Most of them are included in our catalog. \citet{Pedersen2017} presented a careful study on 27 flaring A-type stars. KIC 9216367, KIC 10971633, KIC 11189959 and KIC 12061741 in our sample are confirmed as binary systems by them through radial velocity. They preferred to A-type flaring stars are polluted, but they did not exclude the possibility that they were flaring by themselves. Moreover, even if the A-type star is a binary systems, whether flares can be detected from cool companions is in dispute \citep{Balona2015,Pedersen2017}.

If they have different mechanisms, they may leave clues on their performance such as the statistical character of the profile, but it is difficult to find difference by analysing their properties because of the low time resolution \citep{Yang2018}. Furthermore, the sample of A-type stars in the {\it Kepler} mission is small. A large sample and an elaborate follow-up observation are needed to cope with this issue. We could look forward to the TESS mission that aimed at bright stars by a high time resolution.

 \subsection{Flare Frequency Distribution }
Figure~\ref{figffd} shows FFDs in different spectral types. They obey a power-law relation, i.e., d$N$/d$E$ $\sim\,E^{-\alpha}$. For comparison, we use the same energy range to fit the $\alpha$ index as far as possible. For the solar-like stars, all kinds of flares from nanoflares to superflares follow this relation with $\alpha \sim 2$ \citep{Shibata2013}. It implies that the triggering mechanism of flares are the same, which is derived from the magnetic reconnection. In the {\it Kepler} mission, the results of the late type stars are in accord with the previous works \citep{Maehara2012,Shibayama2013,Yang2017}. Appendix C presents the average cumulative FFDs for each spectral type.

\citet{Audard2000} investigated the $\alpha$ index of 10 stars from F-type to M-type by their EUV flares. They found that for F- and G-type, $\alpha \sim 2.28$, for K-type, $\alpha \sim 1.87$, and for M-type, $\alpha \sim 1.84$. Thus they suggested that there  was a possible dependence of $\alpha$ on the spectral type, i.e., the FFD becomes flat toward later spectral types. In our study, there is a trend from F- to K-type, whereas it turns to steep again in M-type. The $\alpha$ index can be influenced by the energy range that is used to fit, the number of stars, and even the observational moment. It also varies widely for individual star. A general consensus is that it is about 2. Nevertheless, for this trend, we suggest that a significant result and a reasonable explanation are required to support it.

Interestingly, each stellar type has $\alpha \sim 2$ except the A-type star, whose index is much smaller than others. The dashed line in bottom middle panel shows the FFD of fully convective stars, which theoretically should not have the tachocline. The geometry of their magnetic fields are different and may have a switch in the dynamo\citep{Shulyak2017}. Although \citet{Wright2016} argued that they still have an interface dynamo, it is hard to explain the generation of the toroidal field line based on the theory of the Sun\citep[e.g.,][]{Parker1975,Parker1979}. Regardless of whether they have an interface dynamo or other mechanisms such as the $\alpha^2$ dynamo, one thing is certain, i.e., their flares originate from the magnetic reconnection. The similarity of their indices indicates a law that all of the flares caused by the magnetic field follow a power-law relation with $\alpha \sim$ 2.

However, with respect to the A-type star, we even do not know whether it should have a magnetic field without an efficient convection zone. Its $\alpha$ index apparently deviates the common value. Combined with the anomaly of the incidence, it is another indirect evidence that A-type stars generate flares in a different way.

An A-type star has a very weak magnetic field and may have a very thin convection zone \citep{Lig2009}. Nevertheless, it is not surprising that independent X-ray flaring events are observed. In fact, there are many Ae and Be stars \citep{Kogure2007}, whose emission lines are attributed to circumstance such as the accretion disk or the shell. But in the {\it Kepler} mission, many single stars are flaring repeatedly even without emission lines and show large modulation of lightcurves caused by starspots \citep{Balona2013,Pedersen2017}.

Early type stars have convective cores, but according to the solar theory, the magnetic energy cannot flow up without an efficiently convective envelope \citep{Parker1975}. One potential explanation is that magnetic field lines (i.e., poloidal field lines) were frozen in a star when it was born, given that the pre-main sequence stars such as T Tauri stars and Herbig Ae/Be stars often have a strong magnetic field. And the rapid rotation enhances the twisting of those lines. It avoids the issue of generating a magnetic field without a convection zone. However, those large-scale structure field seem too weak to produce such energetic flares.

One caveat is that the situation of the A-type star is complicated. For example, the metallic-line A-type star (Am star) has a much weaker Ca ~{\sc ii} K-line, which is also the indicator of the chromosphere activity. Although it is due to the poor metallicity that is caused by the chemical separation \citep{Char1993,Gray2009}, its strength of the magnetic field is similar to that of normal A-type stars \citep{Blaz2016}. The $\kappa$ mechanism triggers the pulsating process in the early type star \citep{Chevalier1971}. It can ionize plasma in the exterior lays, leading to the variation of the opacity. The change of the opacity could give rise to the change of the luminosity, which may finally result in quasi-profiles of flares. It is not clear whether this process could occur randomly in non-pulsating stars.
\begin{figure}[!htb]
\begin{center}
\subfigure{\includegraphics[width=0.45\textwidth]{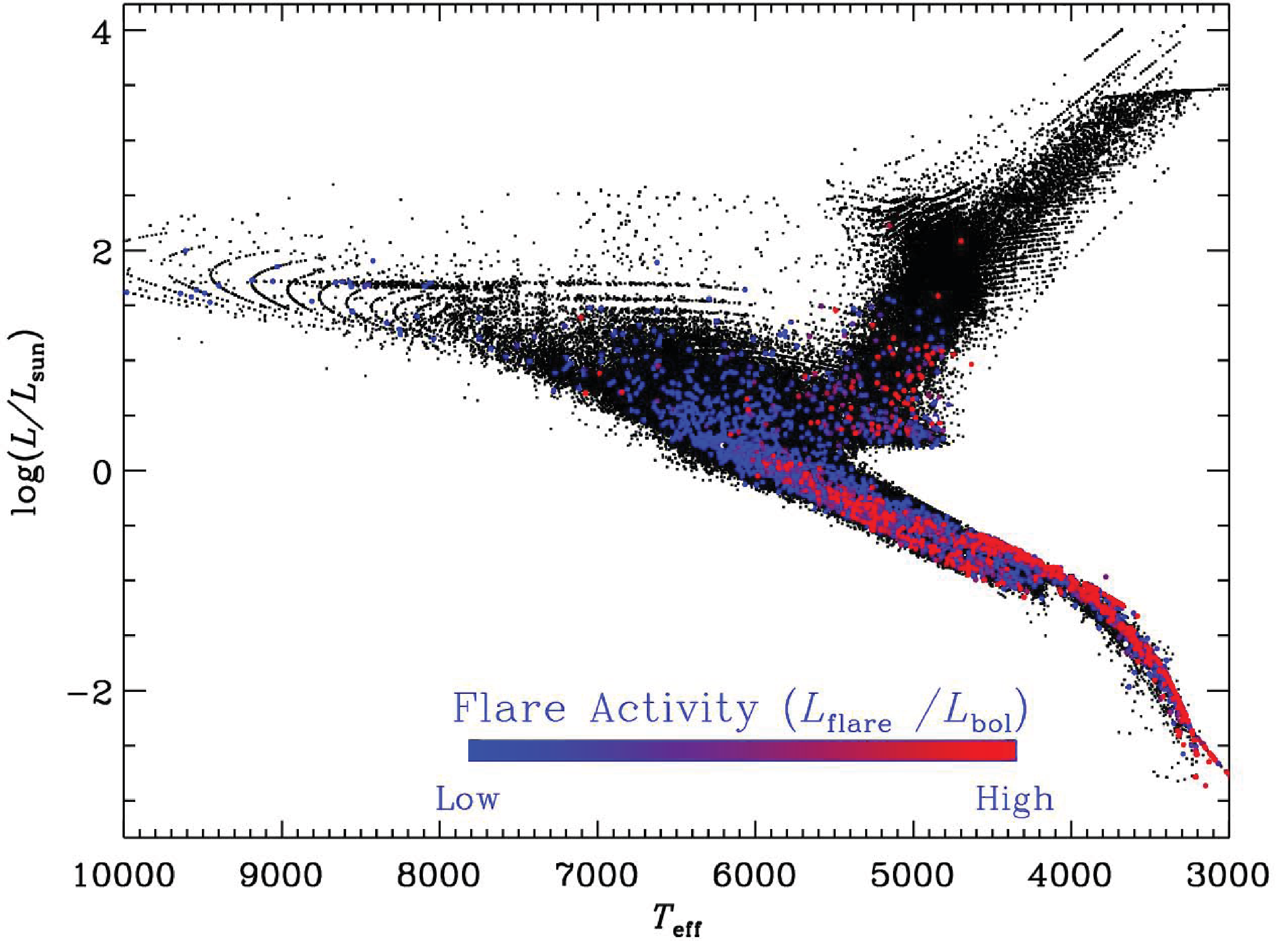}} \hspace{1mm}

\caption{The flare activity across the H-R diagram. Black dots denote over 200,000 Kepler stars. Colorful dots are the 3420 flaring stars. The color changes from blue to red along with the increase of the flare activity. The luminosity of stars are estimated by temperatures and radii given by KIC. Note that the parameters of DR 25 are calibrated by the Dartmouth isochrones \citep{Dotter2008,Huber2014,Mathur2017}, so that one can see regular tracks on the H-R diagram. Some previous versions of KIC may adopt Yansei-Yale isochrones for the calibration.} \label{fighr}
\end{center}
\end{figure}
 \begin{figure*}[!htb]
\begin{center}
\subfigure{\includegraphics[width=0.9\textwidth]{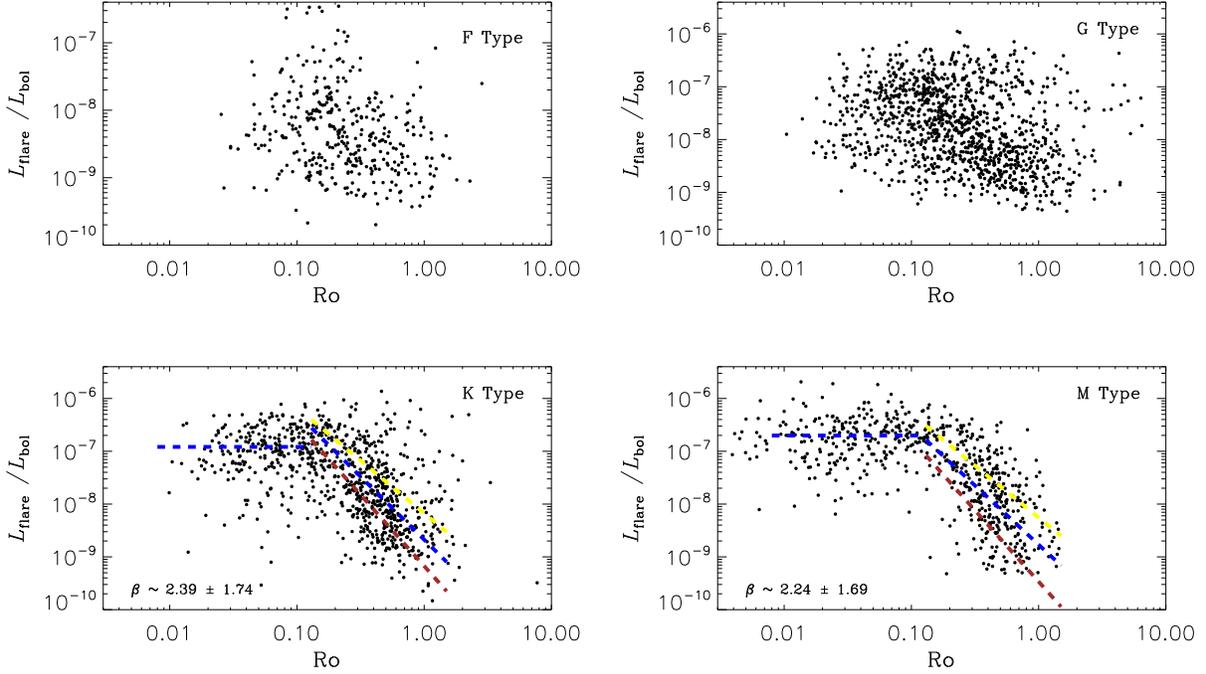}} \hspace{1mm}

\caption{Flare activity vs. Rossby number. The stars are grouped by the spectral type. The $\tau$ of the Rossby number is according to Table 2 of \citet{Wright2011}. Blue lines are obtained by the ordinary least-squares (OLS) bisector fitting in the log-log plane. The error reflects the uncertainty of the parameter whose deviation of $\chi^2$ from its best-fit value reaches $\Delta\chi^2$ = 1. The turnoff point is Ro=0.13 as suggested by \citet{Wright2011}. The index $\beta$ indicates the slope of the unsaturated regime of the blue line. Red lines and yellow lines are with $\beta = 2.7$ and $\beta = 2.0$ respectively for comparison, which have been vertically shifted for clarity.} \label{figactao}
\end{center}
\end{figure*}

 \begin{figure*}[!htb]
\begin{center}
\subfigure{\includegraphics[width=0.9\textwidth]{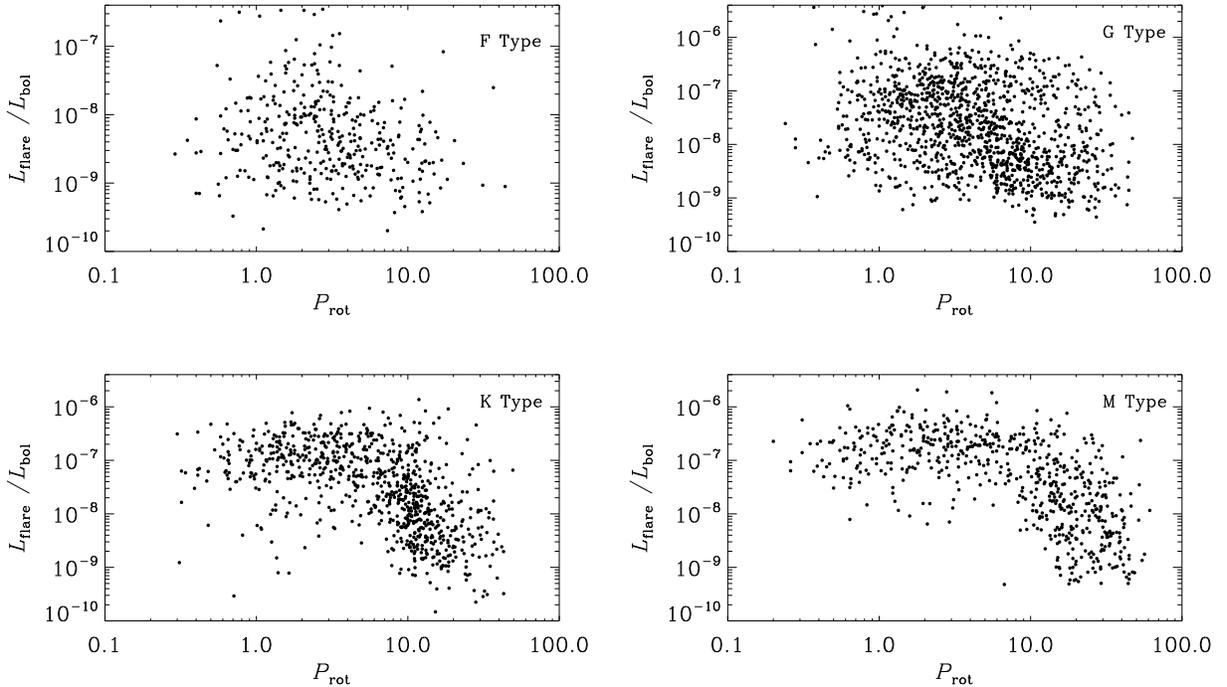}} \hspace{1mm}

\caption{The same as Figure~\ref{figactao}, but for the rotation period.} \label{figactper}
\end{center}
\end{figure*}
  \begin{figure*}[!htb]
\begin{center}
\subfigure{\includegraphics[width=0.95\textwidth]{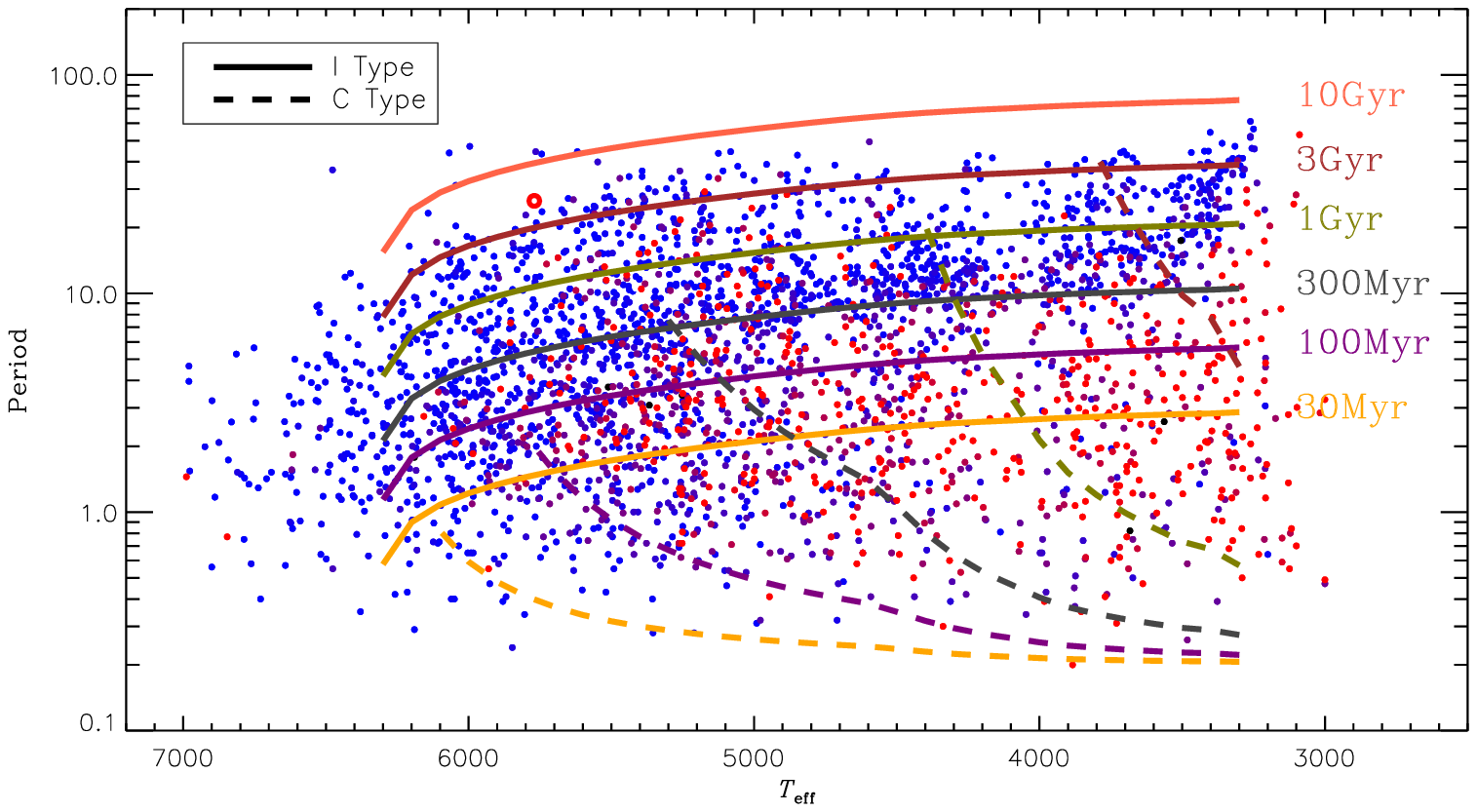}} \hspace{1mm}

\caption{Rotation period vs. temperature. The Lines are the empirical ischrones of Gyrochronology, of which the solid line and dashed line represents the I sequence and the C sequence respectively. The same color of the ischrones denotes the same age. The bifurcations of the lines indicate the transition point when C stars evolve onto I stars. The Flare stars are overplotted in the diagram by filled circles. The color of the circle represents the flare activity, which is as the same as Figure~\ref{fighr}. The Sun is marked as the open circle. The ischrones of the I sequence is given by \citet{Mamajek2008}, while the ischrones of the C sequence is according to \citet{Barnes2010}. The differences of ischrones of the I sequence among \citet{Mamajek2008}, \citet{Meibom2009} and \citet{Barnes2010} are too small to affect the analysis and scenario of this study. Transformations from color to temperature are accomplished using Table 1 of \citet{BarKim2010}.} \label{figage}
\end{center}
\end{figure*}
\section{The Flare Activity}

Figure~\ref{fighr} shows the flare activity across the H-R diagram. Colorful dots are 3420 flaring stars over 200,000 background stars in the {\it Kepler} mission. The flare activities are listed in Table 1. Their accuracy could be influenced by the profile of LC data and the lost flares. As we mentioned in Section 2.4, we have calibrated the underestimation of LC energies. And the lost flares have little influence on the flare activity.

It is clear that active stars distribute at the late type region. 6000 K is a dividing line for the flare activity, given that most active stars are on its right side. Some giants show high activities, which go against the knowledge on the dynamo and the evolution. We will address this issue in Section 4.5. A few stars are unclassified and their parameters are set as the same as the Sun \citep{Mathur2017}. Those stars are not in our following studies.

 A reliable and large sample of the flare stars provides us a chance to gauge the present theories of stellar activity. In the next subsections, We will focus on the issues among the flare activity, the rotation period, the age and the dynamo by reviewing previous studies and exploring their relations.

 \subsection{The Activity--Rotation Relation}
The indicator of the stellar activity in the X-ray band and the chromosphere band is $R_{\rm X} = L_{\rm X}/L_{\rm bol}$ and $R'_{\rm HK} = L_{\rm HK}/L_{\rm bol}$ respectively. Although they are from different bands and represent different stellar layers, they have the same form on the function of the rotation, consolidating the activity--rotation relation.

The flare activity is defined as $R_{\rm flare} =\sum E_{\rm flare}/\int L_{\rm bol}dt = L_{\rm flare}/L_{\rm bol}$. The Rossby number is the ratio of the rotation period to the convective turnover time, i.e., Ro$ = P_{\rm rot}/\tau$. The $\tau$ reflects the depth of the convective envelope and the convective velocity therein, which can be obtained by a theoretical calculation or an empirical fitting \citep[e.g.,][]{Noyes1984,Pizzolato2003,Wright2011}. It relates the theoretical dynamo to the observation, because a strong support is that the activity--rotation relation will become tight by introducing $\tau$.

Figure~\ref{figactao} shows $R_{\rm flare}$ as a function of Ro on different spectra types. For K- and M-type stars, the flare activity increases with decreasing rotation period, demonstrating a classical activity--rotation relation, i.e., the saturation regime and the exponential decay regime corresponds to fast rotation stars and slow rotation stars respectively. The saturation regime is usually explained by  the magnetic saturation \citep{Reiners2009}. The switch between the saturation and the unsaturation can be attributed to the geometry of the magnetic field that caused by the modulation of the dynamo \citep{Shulyak2017}. The slope of the decay is $\beta \sim 2.39 \pm 1.74$ where rms = 0.72 dex and $\beta \sim 2.24 \pm 1.69$ where rms = 0.67 dex respectively given by the best fitting model of the ordinary least-squares (OLS) bisector in the log-log plane \citep{Isobe1990}, which is consistent with most of the previous studies \citep[e.g.,][]{Noyes1984,Pizzolato2003}. The error reflects the uncertainty of the parameter whose deviation of $\chi^2$ from its best-fit value reaches $\Delta\chi^2$ = 1. While it may vary between 2.0 and 2.7 \citep{Gudel1997,Wright2011}, we also plot the lines with $\beta \sim 2.7$ and $\beta \sim 2.0$ for comparison. Those lines are vertically shifted for clarity. Figure~\ref{figactper} is as the same as Figure~\ref{figactao} but for the rotation period, showing that the dispersion is larger. \citet{Davenport2016} presented the relative flare luminosity as a function of Rossby number for stars later than G8, which is similar with Figure \ref{figactao}. However, in his result, the critical Rossby number separating the saturated and decay regimes is $\sim 0.03$ and $\beta \sim 1$, which are much different with that found in the X-ray band. As we have mentioned in Section 3.1, the data he used have a high contamination rate, which could seriously affect the analysis.

As shown in Figure~\ref{figactao}, the activity--rotation relation gradually becomes dispersive with increasing temperature, and nearly disappear in G- and F-type stars. This phenomenon is also observed in the X-ray band (e.g., see Figure 8 of \citet{Pizzolato2003}), indicating that the scaling factor of the activity--rotation relation becomes sensitive to increasing temperature. The trend is not reported by the indicators of $R'_{\rm HK}$, because they cannot be measured in low mass stars.

One intriguing question is what is the nature behind this trend, since the Rossby number is considered as the unique variable in stellar activity. We will address this issue by invoking the age--rotation relation: Gyrochronology.

\subsection{Activity,Rotation and Age}
\begin{figure*}[!htb]
\begin{center}
\subfigure{\includegraphics[width=0.90\textwidth]{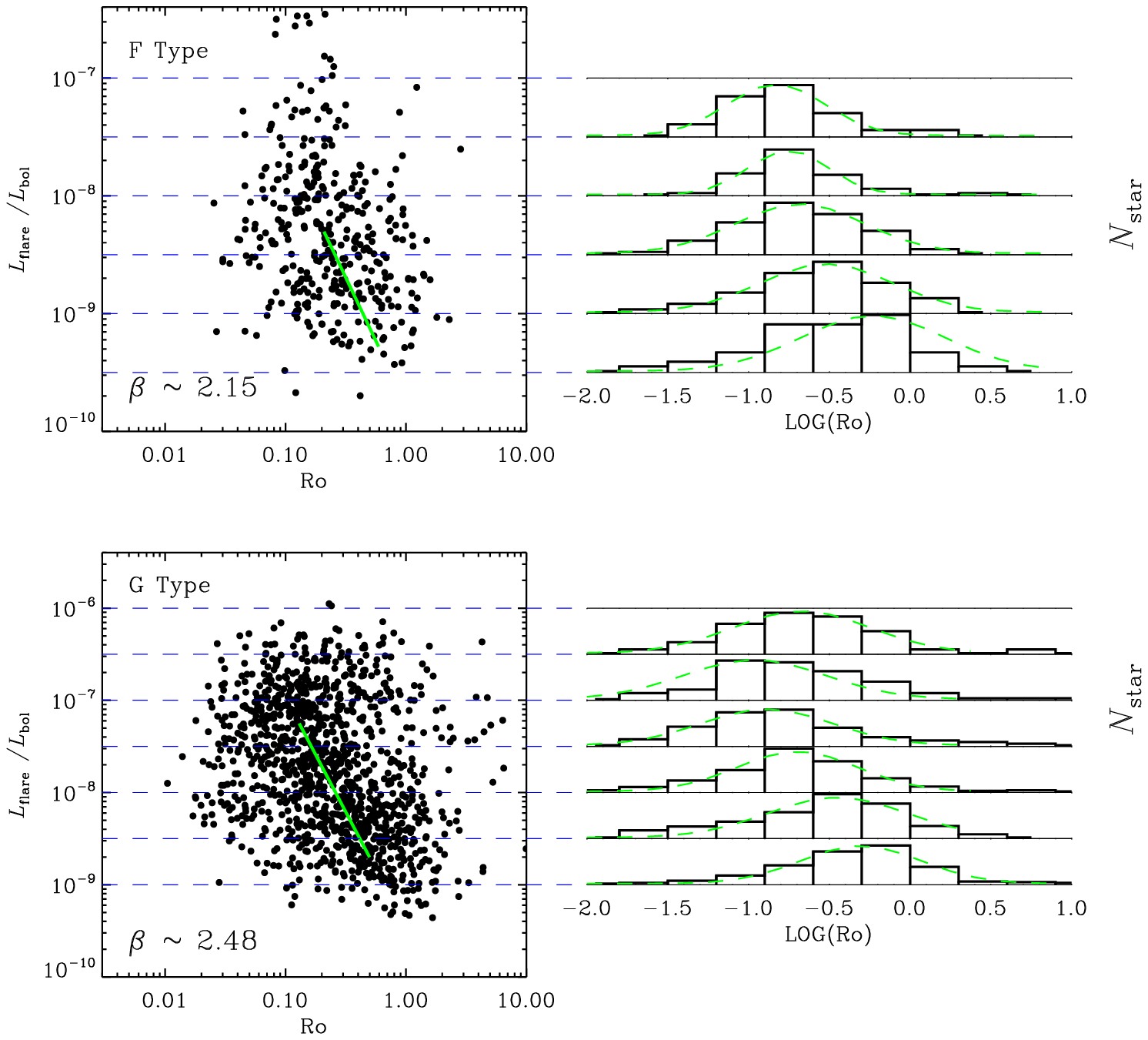}} \hspace{1mm}

\caption{ A detailed analysis on the activity-rotation relation for G type and F type stars. It shows that the saturation regime has disappeared and the unsaturation regime still follows a law of exponential decay, indicating that C stars become sensitive to temperature, while I stars always follow the same law. Blue dashed lines of left panels separate flare stars into bins by the flare activity. In each bin, the distribution of flare stars along the Rossby number is plotted in right panels. Each distribution is fitted by a gaussian function that is plotted by a green dashed line. Centers of some gaussian functions and the average activities of corresponding bins are applied to fit the activity-rotation relation, which is plotted as the green solid line in the left panel. Its slope is marked as $\beta$.} \label{figcd}
\end{center}
\end{figure*}
The age--rotation relation is aimed at predicting the age by the rotation period for the main sequence (MS) star, which is also dubbed as Gyrochronology \citep{Barnes2003,Barnes2007,Barnes2010}. Inspired by the distribution of rotation periods in open clusters, \citet{Barnes2003} proposed that the MS stars can be grouped by two different stellar dynamos, which are coined as convective (C) sequence and interface (I) sequence.

The I sequence stars (hereafter I stars) are slow rotators and have the same dynamo with the Sun. They lose angular momentum by the interaction between the open field lines and the ejective materials, i.e., magnetized stellar wind \citep[e.g.,][]{Gallet2013}, which builds a connection between the rotation and the age. The C sequence stars (hereafter C stars) are young and fast rotators with a convective dynamo that is not clearly understood yet. \citet{Barnes2003} pointed out that the C star had two decoupled parts, i.e., a radiative core and a convective envelop, when it was born. As it is aging, the two parts gradually become coupled to each other. Finally C stars evolve onto the I sequence and their dynamos switch to the solar-like dynamo. Considering the causality and outliers between the I sequence and the C sequence, \citet{Barnes2003} took the gap between the two sequences as the transition sequence.

The outline on C stars is a little ambiguous, because it does not explain why there are two naturally separated parts at the beginning. In spite of the dynamo,  the morphology of the period--color diagram indeed indicates the two sequences and their relations. The Gyrochronology has greatly developed the accuracy on the age estimation for main sequence stars compared with the traditional isochrones, although its errors may be larger after 2.5 Gyr \citep{Vans2016}. Note that a fully convective star does not have a radiative core because of its mass. C stars definitely differ from it, which may give rise to misunderstanding.

The rotation, the activity and the age can relate each other. One can combine them to give a calibration \citep[e.g.,][]{Mamajek2008}. By comparing the activity--rotation relation and the age--rotation relation, \citet{Barnes2003b} associated I stars with unsaturated stars (in the exponential decay regime) of the activity--rotation relation, as both of the two populations have relatively long periods. Equally, he associated C stars and the gap with saturated stars. The morphology of the two relations indeed demonstrate the correspondence, because the boundary of the I sequence and the C sequence is nearly as the same as the boundary of the saturation regime and the unsaturation regime, i.e., the turnoff point of the activity--rotation relation can also separate the I and the C sequence. Moreover, the simplicity on interpreting the turnoff as two different dynamos based on different evolution stages makes it more appealing than resorting to the modulation of the same dynamo.

Figure~\ref{figage} shows the empirical isochrones of the Gyrochronology \citep{Mamajek2008,Barnes2010}, on which the 3420 flaring stars are superimposed. The isochrones use F5 ($B-V \sim 0.47$) as an anchor point of the $x$ axis and young stars of open clusters and the Sun as anchor points of the $y$ axis. The lines with the same color are the coeval isochrones. The bifurcations of the isochrones represent transition points from the C sequence to the I sequence. The function of the I sequence is extensively discussed and calibrated \citep [e.g.,][]{Mamajek2008,Meibom2009,Garca2014}. Nevertheless, it should be noted that the period of the C star is described by $P_C = P_0e^{t/T(B-V)}$. It strongly depends on the initial period $P_0$ and the relative spin-down timescale $T(B-V)$. $T(B-V) = \tau/k_C$, where $k_C$ is a dimensionless constant that is obtained by statistics of observations in open clusters \citep{Meibom2009}, whereas $k_C$ only has 4 anchor points, i.e., only 4 points are applied to fit $T(B-V)$ from the G-type to the M-type. We set the initial period of the C sequence $P_0=0.2$ following \citet{Barnes2010}.

As shown in Figure~\ref{figage}, the isochrones of the two sequences are separated at the low temperature band and converge at the high temperature band. In the activity-rotation diagram, this trend makes the two sequences naturally separated by the rotation period in K- and M-type stars, resulting in the saturation regime and the unsaturation regime. But for the F- and G-type stars, the loss rates of the angular momentum of C stars rise rapidly. The rotation period cannot distinguish C and I stars, because they gradually mix together when the temperature increases. The convergence also leaves hints on the range of the rotation period. Figure~\ref{figactper} shows the range of the rotation period gradually shrink towards to the early type star.

This scenario predicts the disappearance of the swerve on the early type stars. And the disappearance of the swerve is proved by Figure~\ref{figactao}. However, if C stars correspond to a unchanged activity (the saturation value) and I stars follow a law of exponential decay, their activity level should be different as before. The shape of their relation should be similar to the letter `T', i.e., the two sequences still can be separated by the flare activity. But we do not see the separation in the F-type and the G-type stars.

As shown in Figure~\ref{figcd}, in the light of the flare activity, we divide the flaring stars into bins, then measure the distribution of the flare number in each bin. In the right panels, a Gaussian function (the green dashed line) is applied to fit the number distribution in each bin. The center of each distribution represents the overlap of C and I stars and it shifts along with Rossby number. Green lines of left panels are fitted by the central shift and the average activity. Their slopes are $\beta \sim $2, indicating that I stars still follow a law of exponential decay, whereas C stars scatter everywhere.

Figure~\ref{figcd} indicates that the relation between I stars and C stars is different from the foregoing relation. With respect to the I sequence stars, they remain a power-law relation, which is reflected by the shift of the distribution. Because of the scatter of C stars, nor can the activity level separate the two sequences. The scatter demonstrates that the activity of C stars becomes sensitive to temperature. The saturation level attains a constant value in K- and M-type stars, and varies in G- and F-type stars, implying that the stellar dynamo of the C sequence has a non-linear relation with the temperature.

The overall picture of the activity--rotation relation is presented here. It has a dynamic variation across the H-R diagram and this variation inspires us to reassess the activity--rotation relation from the dynamos and their correspondence to the activities and the rotations.

\subsection{The Activity of The C Sequence Stars}
\begin{figure}[!htb]
\begin{center}
\subfigure{\includegraphics[width=0.5\textwidth]{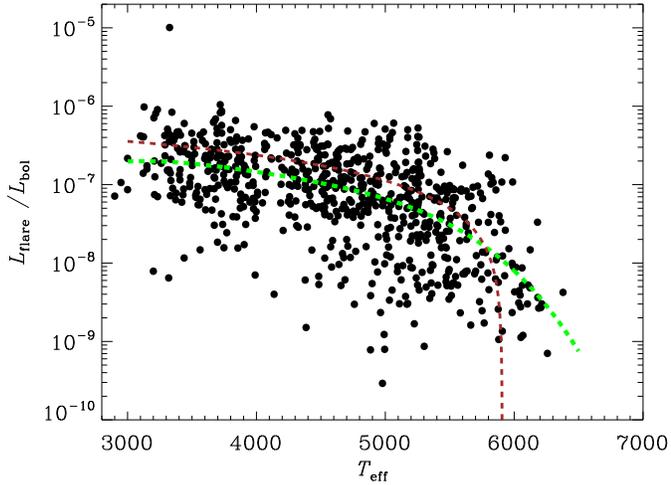}} \hspace{1mm}

\caption{The flare activity of C stars as a function of the temperature. The Figure comprise stars whose rotation period is shorter than the corresponding period of the 30 Myr ischrone of the I sequence. The temperature range is $3000 <T_{\rm eff} < 6400$. Most of those stars are C stars, although a small part of them may be I stars that are younger than 30 Myr. The green dashed line is a polynomial fitting and the red dashed line is a gaussian fitting.} \label{figca}
\end{center}
\end{figure}

\begin{figure*}[!htb]
\begin{center}
\subfigure{\includegraphics[width=0.90\textwidth]{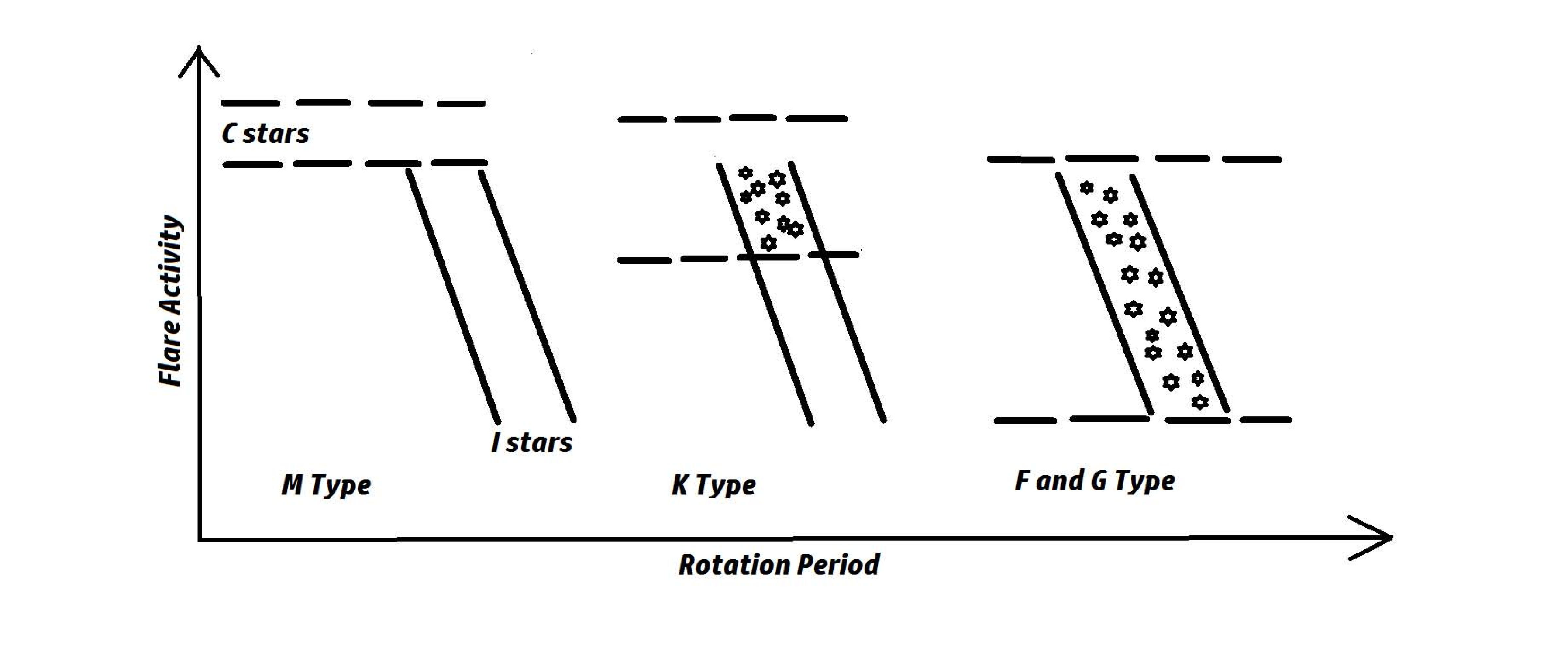}} \hspace{1mm}

\caption{Sketch map of the activity-rotation relation on C and I stars. Dashed lines denote the region of C stars and solid lines denote the region of I stars. The overlaps of C and I stars are filled with asterisks. I stars always keep a power-law relation on the Rossby number. C stars of the M-type are fast rotators and less sensitive to the temperature. They are separated from I stars in different regimes. As the temperature increases, rotation periods of C stars begin to converge with I stars and the activities of C stars become sensitive to the temperature resulting in the mixing of C and I stars both in the rotation period and the activity. } \label{figskt}
\end{center}
\end{figure*}
In the X-ray band, the variation of the saturation level can be seen in Figure 7 of \citet{Pizzolato2003} as well. The sample of \citet{{Pizzolato2003}} comprise 259 dwarfs ranging from F8 to M0, which shows that the $(L_{\rm x}/L_{\rm bol})_{sat}$ drops rapidly near the stellar type of the Sun. \cite{Wright2011} used a much larger sample (824 stars), in which about 20\% stars are earlier than K0. Although it includes the sample of \citet{Pizzolato2003}, they applied a uniform fitting to all of the stars, and thereby do not report the variation of the saturation value.

With respect to $R'_{\rm HK}$, most stars of those samples are solar-like stars, and the activity--age relation requires a distinct conclusion on the color dependence of $R'_{\rm HK}$. The activity of I stars have an association with the color through the convective turnover time $\tau$, and $\tau$ is not the unique factor, resulting in a weak color dependence \citep{Mamajek2008}. The solar-like stars of the C sequence have a strong dependence on the color. Figure 2 of \citet{Noyes1984} indicates that those `young' stars have a similar trend to Figure 7 of \citet{Pizzolato2003}, while their activities are almost not affected by the rotation period (see Figure 3 of \citet{Noyes1984}). \citet{Soder1991} found that the halo stars(I stars) have no dependence on color, while Hyades cluster has a clear relation, whose member consist of 10\% C stars \citep{Meibom2009}.

Figure~\ref{figca} shows the color dependence of C stars, It includes stars whose rotation periods are shorter than the corresponding periods of 30 Myr ischrones of the I sequence, i.e., their regime is below this ischrone as shown in Figure~\ref{figage}. This regime minimizes the pollution from the I sequence. Most of them are C stars, although a small part of them could be very young I stars. Above this isochrone, the components of those regions become heterogeneous so that the two sequences cannot be separated.

Previous studies acquiescently take the saturation regime as a constant, especially for those low temperature stars. However, given that the two regimes correspond to different dynamos rather than the modulation of a dynamo, a more appealing interpretation is that the activity of C stars is continuously changing along the color in a non-linear form. Figure~\ref{figca} shows this trend. The activity of the C star is less sensitive to the color at the low temperature band and more sensitive to the color at the high temperature band. A quadratic polynomial fitting of the trend given in Figure~\ref{figca} is

\begin{center}
\begin{equation}
{\rm log}\frac{L_{\rm flare}}{L_{\rm bol }} = -9.20 + 1.41\times 10^{-3}T_{\rm eff}-2.03\times 10^{-7}T_{\rm eff}^2
\end{equation}
\end{center}
or
\begin{center}
\begin{equation}
{\rm log}\frac{L_{\rm flare}}{L_{\rm bol }} = -10.03 + 4.49(B-V)-1.56(B-V)^2.
\end{equation}
\end{center}

The expression demonstrates that the temperature is the unique factor for the activity of C stars. As known to the solar-like dynamo, magnetic filed lines are generated and strengthen in the tachocline by interaction between a core and an envelope \citep[e.g.,][]{Schou1998}. The shear between the two layers relates to the rotation and the differential rotation. Nevertheless, If the dynamo of C stars is from the interior of the convection zone, it is reasonable that the magnetic activity does not depend on the rotation and the differential rotation. An appropriate assumption is that it still associates with the convective turnover time but in a different and non-linear form, which leads to the dependence on the temperature.
 \begin{figure*}[!htb]
\center
\subfigure{\includegraphics[width=0.9\textwidth]{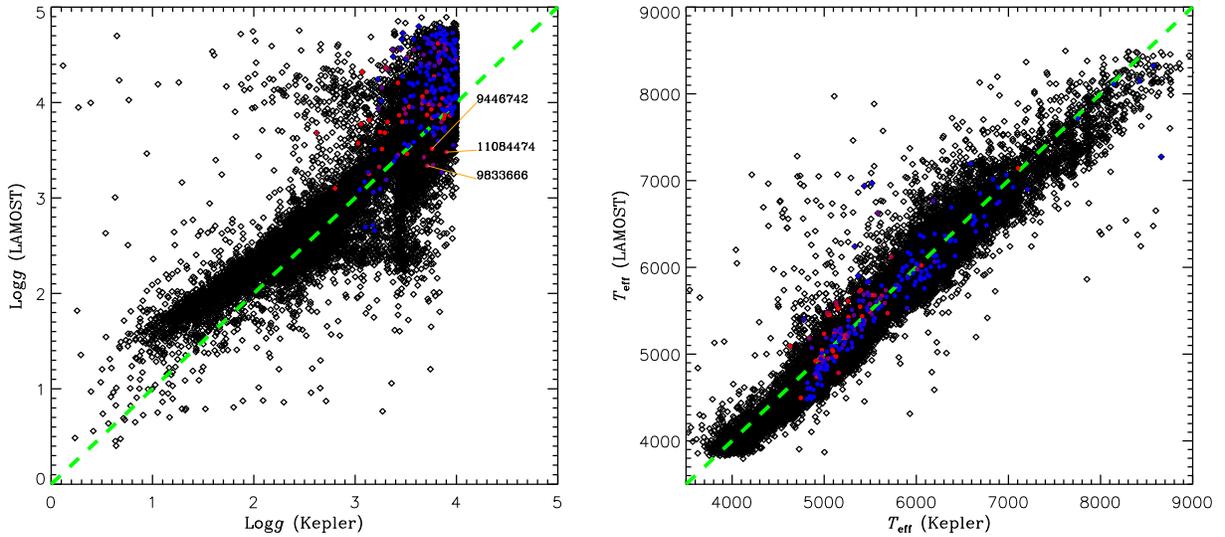}}
\hspace{2 mm}
\caption{The comparison of the log$g$ (left panel) and the temperature (right panel) between {\it Kepler} and LAMOST. The black diamonds are the 25,576 {\it Kepler} targets with the KIC parameter log$g < 4$. The colorful circles are the 221 flaring giants. The color represents the flare activity, which is as the same as Figure~\ref{fighr}. The green dashed lines are the equivalent lines. } \label{figgia}
\end{figure*}

 \subsection{The Expression of the Flare Activity}

In the above sections, we investigate the relation between activity--rotation diagram and the Gyrochronology. The scenario is illustrated by Figure~\ref{figskt}. The C dynamo depends on the temperature rather than the rotation period, while the I dynamo depends on the Rossby number. The C and I dynamo generate different activities. As they have distinctly different rotation periods in K- and M-type stars, their activity regimes are naturally separated. As the temperature increases, the C dynamo becomes sensitive to the temperature. Meanwhile, its loss rate of the angular momentum rise rapidly, resulting in a longer rotation period. Those transitions make the two population stars mix together in the activity--rotation diagram. The overlap of the two sequences causes a higher density of the star number, which is shown in Figure~\ref{figcd}. The I sequence follows a law of the exponential decay so that the shift of the density can be seen.

The activity--rotation relation is not available for C sequence stars. To summarize the activity--rotation--color relation in a uniform expression, we suggest its form to be

\begin{center}
\begin{equation}
{\rm log} R_{\rm act} = \begin{cases} C_0+C_1T_{\rm eff}+C_2T_{\rm eff}^2,&{\rm C~~Sequence} \cr -2{\rm log R_o}+C, &{\rm I~~Sequnece}\end{cases}.
\end{equation}
\end{center}
The scaling factors $C$ and $C_i$ are different for different bands.

We should note that the flare activity contains the parameter $L_{\rm bol}$ that relates to temperature and $\tau$ so that the relation between the activity and the parameters are intrinsically not independent. Strictly speaking, whatever dynamo a star has, its activity depends on temperature. However, the normalized activity is taken as a whole quantity in the analysis to present its sensitivity to temperature for different dynamos. The result demonstrates that the normalized activity of the C dynamo is sensitive to temperature but that of the I dynamo is not. Or in another word, the C dynamo has an additional sensitivity to temperature compared to the general relation.

Although the activity of the C sequence star does not depend on the rotation, the rotation restrict the age of a C star. If the rotation period of a star exceeds a threshold, it will be taken as a I star. However, As shown in Figure~\ref{figage}, some solar-like stars have relatively high activities, even if when $P_{\rm rot} > 10$ days. It is possible that some C stars remain their capacity of generating a high activity when they have evolved onto the I sequence. Because the stellar wind is due to open magnetic field lines, while the stellar activity is related to closed magnetic field lines \citep[e.g.,][]{Holzwarth2007}. After a C star evolves onto the I sequence, its loss rate of the angular momentum is consistent with I stars, while their activities are different.

This explanation can resolve a discrepancy in the activity--age relation. For example, Figure 4 of \citet{Mamajek2008} shows the color dependence of $R'_{\rm HK}$ for different clusters. One can see for a given color of the same cluster, there are many outliers, i.e., two stars with the same age and color may have much different activities. It challenges the activity--age relation and is hard to explain by the same dynamo. This assumption also can explain the different activities of a binary \citep{Donahue1998,Mamajek2008}.

 \subsection{The Flare Activity of Giants}

\citet{Door2017} found over 600 flaring giants in quarter 15 of {\it Kepler} data, whose flaring incidence is similar to the main sequence star (see Table 3). However, as we mentioned above, most of them are pulsating stars. In our sample, 76 flaring stars are with log$g < 3.5$, and 358 flaring stars are with log$g < 4$. Although the incidence is much smaller than that of the main sequence star, some giants or subgiants show relatively high activities (see Figure~\ref{fighr}).

According to the dynamo theory and the observation, magnetic field of giants should be weak \citep[e.g.,][]{Simon1989,Kon2008}. If the giant has the same mechanism on generating stellar activity, active giants completely deviate from the activity--age relation. Moreover, as the giant has a much brighter luminosity, its flare energies should be much higher so that they can be detected.

The surface gravity of flaring giants are underestimated in the KIC. Figure~\ref{figgia} shows a comparison of the log$g$ and the temperature between {\it Kepler} and LAMOST. LAMOST is a spectral survey that fully covers the {\it Kepler} field \citep{Cui2012}. The LAMOST--{\it Kepler} project has provided spectra of more than 76,000 {\it Kepler} targets \citep{Decat2015,Zong2018}. There are 25,576 {\it Kepler} targets whose log$g < 4$ having spectral parameters \citep{Luo2015}, of which 221 targets are flaring stars.

The left panel of Figure~\ref{figgia} shows that KIC has systematically underestimated the log$g$, while the right panel shows that their temperatures are basically in consistency with each other. Systematical errors could exist in two independent systems, but the difference of log$g$ between LAMOST and the Sloan Digital Sky Survey (SDSS) is $0.16 \pm 0.22$ dex, which is much smaller than that shown in the left panel \citep{Luo2015}. Many flaring giants are underestimated by over 0.5 dex, which actually are main sequence stars. The estimation of the flare energy depends on the surface gravity. The difference of 0.5 dex will give rise to the deviation of the flare activity by an order of magnitude, suggesting that most giants and subgiants are inactive.

Giants are repeatedly and radially expanding and shrinking. This process involves violently material movements. It is possible to cause a shock between different layers resulting in a false positive signal. There are several stars that are overestimated on the log$g$. Some inactive outliers may have false positive signals or persist their magnetic activities \citep{Harper2013}. For example, Strong flares on a red-clump giant are reported, although it is a small subset of the giant star population \citep{Ayres2001}. However, three active giants are marked in the left panel of Figure~\ref{figgia}. They are frequently faring and their lightcurves have no difference with those of normally flaring stars. We should note that although strict checks have been applied to prevent pollution, we cannot completely exclude the possibility that they have very active companions.

In the cross sample, 81 (36.7\%) flare stars are bona-fide giants with log$g$ $<$ 4 and only 19 of them are with log$g$ $<$ 3.5, indicating that most of them are sub-giants. It is reminiscent of RS CVn binary system, which has a G-type main sequence star and an evolved F-type subgiant \citep{Stra1993}. As shown in the right panel of Figure~\ref{figgia}, the temperatures of the flaring giants are consistent with RS CVn type stars. This binary system has higher activity on its primary (i.e., the main sequence star), and its activity depends on the binary interaction (see \citet{Kogure2007} and references therein).

From the view of flare stars, we suggest that besides a fraction of the anomalies, most giants are inactive, which is in consistent with the dynamo theory and the magnetic observation.

\subsection{The Rotation distribution of Flare stars}

\begin{figure}[!htb]
\begin{center}
\subfigure{\includegraphics[width=0.5\textwidth]{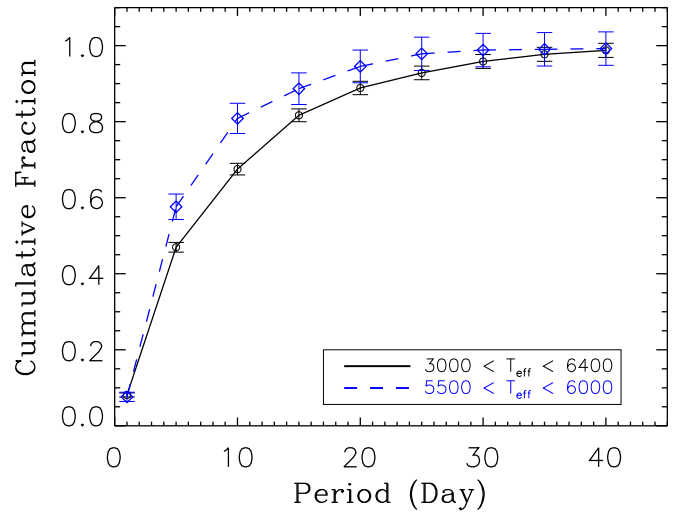}} \hspace{1mm}

\caption{The rotation distribution of flare stars. The $y$ axis is the cumulative fraction of flare stars whose rotation period are shorter than corresponding value given by the $x$ axis. The solid line represents all of flare stars with 3000 K $< T_{\rm eff} <$ 6400 K, while the dashed line represents stars with 5500 K $<T_{\rm eff} <$ 6000 K. The uncertainties in fractions are estimated by the square root of the number of stars.} \label{figpor}
\end{center}
\end{figure}
Since \citet{Skumanich1972} formally proposed that stellar rotation decreased with increasing age, the stellar rotation, as the indicator of stellar evolution or stellar age, has been extensively discussed. For flare stars, their rotation distribution could reflect their age distribution and the variation of their activities along with evolution. It could also provide a reference of the superflare frequency on our Sun.


Figure~\ref{figpor} shows the rotation distribution of flare stars in the {\it Kepler} mission. The $y$ axis is the cumulative fraction, which is calculated by

\begin{center}
\begin{equation}
F_{\rm cum}(P_0) = \frac{N(P<P_0)}{N_{\rm total}}.
\end{equation}
\end{center}
$N(P<P_0)$ is the number of flare stars whose rotation period are shorter than a given period $P_0$. $N_{\rm total}$ is the total number of flare stars. The solid line is all of the applicable stars with 3000K $< T_{\rm eff} <$ 6400 K and the dashed line represents stars with 5500 K $<T_{\rm eff} <$ 6000 K.

The cumulative fraction attains 50\% when the rotation period is near 6 days that is also the turn off point of the activity--rotation relation. It implies that flare stars on the C and I dynamo have the similar number. About 70\% flare stars rotate faster than 10 days and the fraction approaches 95\% at 30 days. According to Figure~\ref{figage}, their ages are younger than 1 Gyr and 3 Gyr respectively. At most 5\% flaring solar-like stars rotate slower than 25 days by taking into account the error.

Strictly speaking, a solar-like star should have many constraints such as mass, age, metallicity and even its circumstellar material. Nevertheless, from the view of stellar structure and evolution, the most important factors are its mass and age, which can be represented by the effective temperature and rotation period to a certain extent. We thus take stars that are similar with the Sun on the two parameters as solar-like stars and estimate the superflare frequency on them.

The most famous solar flare is the Carrington event, whereas there is no accurate estimation on its energy. The largest solar flares that have strict energy estimation on the Sun is of order $10^{32}$ erg \citep{Emslie2012,Moore2014}. Considering that the Carrington event is the largest proton flare event during the past 450 yr and its prominent contribution to the white light band \citep{Shea2006}, it is possible that the Carrington event has a higher order of energy than $10^{32}$ erg. However, even such flare energy may not be observed by the {\it Kepler} mission because of the detection limit. As shown in Figure~\ref{figffd}, only if a solar-like star produce a flare with energy $\sim 10^{34}$, it will be certainly detected by the {\it Kepler} mission. We hence assume the flaring solar-like stars in our sample can at least produce a such energetic flare in 4 years. The incidence of G-type flare stars is 1.46\% according to Table 3. Given the observation time of the {\it Kepler} mission, a solar-like star with $P_{\rm rot} > 25$ days has about $1.46\% \times 5\% \sim $ 0.07\% probability to generate a flare with energy $\sim 10^{34}$ erg in 4 years, implying that such flare at least occurs on the Sun once in $\sim$ 5500 years.

\citet{Shibayama2013} estimate that a superflare with energy $\sim 10^{34}$ erg occur on a solar-like star (5600 K $< T_{\rm eff} <$ 6000 K, $P_{\rm rot} >$ 25) once in $\sim 2000$ years. In their estimation, they calculate the mean frequency of superflares in solar-like stars. However, due to a quite small sample, this frequency could be seriously influenced by individual active star. Thus we assume all of the flaring solar-like stars have the same capacity, i.e., they can produce at least a superflare in 4 years. It gives the lower limit of the superflare frequency on the Sun.

\subsection{Flares and Starspots}

 The lightcurve of the {\it Kepler} mission shows that starspots often concentrate near a longitude. Its amplitude could be a proxy for the size of the starspot \citep{Cand2014}. If one looks up lightcurves of the {\it Kepler} mission, there are many stars having large starspots but no flares. In fact, over 34,000 main sequence stars show modulations of lightcurves that could be caused by starspots \citep{Mc2014}. In those stars, about 2500 (7\%) stars have flares and there is no correlation of flare occurrence or energy with starspot phase \citep[e.g.,][]{Hawley2014,Doyle2018}. \citet{Yang2017} investigated the relation between the flare activity and the amplitude of starspots, finding that they have a positive correlation and this correlation was influenced by the inclination angle.

This correlation is reasonable, given that the starspot is the manifestation of the magnetic flux tube and the magnetic reconnection is related to the interaction of a pair of starspots. However, this correlation is merely available for 7\% of stars and 93\% of stars do not produce energetic flares, although they have large starspots. One possible explanation is that it depends on the magnetic configuration or the topology of the magnetic field. However, the remarkable proportion makes it implausible that the starspot directly relates to the flare activity. If we switch to another point of view, one doubt is that are all of the modulations of lightcurves caused by starspots? There are few alternative options because the number of ellipsoidal binaries in the {\it Kepler} mission is small \citep{Gao2016}.

On the other hand, 291 stars of our sample do not show apparent variations of lightcurves. \citet{Yang2017} also reported this phenomenon. Some of them are very active. One reason could be due to the inclination angle of the stellar rotation with respect to the observer's line of sight. However, It seems that the flare will also be influenced by the inclination angle, given that a flare occurs between a pair of starspots.

Another explanation could be that starspots concentrate near the rotation pole so that their capacity on the modulations of lightcurves will be much lower than that when they are near the equator. Fast rotators with high activities could have polar spots \citep{Sch1992}, whereas many stars of our sample are inactive. A more acceptable scenario requires a new dynamo. In fact, if there is a fine tuning of the $\alpha-\Omega$ dynamo by assuming a new mechanism on the generation of the poloidal field and the meridional circulations of stars are weak or have an opposite direction with the Sun, polar spots will appear \citep{Wang1991,Choudhuri1995}.


The different layers should be taken into account as well. The starspot is in the photosphere, and the flare occurs in the chromosphere. Their relation could shed light on the structure and the energy transportation in those stars.

\section{Conclusions}
We search flare events through LC data of DR25 in the {\it Kepler} mission and present 3420 flaring stars in over 200,000 {\it Kepler} targets by excluding various false positive signals and artifacts. The number of flare events is 162,262. The catalog has improved dramatically on previous catalogs \citep{Davenport2016,Door2017}, which exhibit a lot of contamination (The contamination rates are over 60\%).

The incidence of flare stars from the F-type to the M-type increases from 0.69\% to 9.74\%. The result is qualitatively consistent with the general framework of $\alpha-\Omega$ dynamo, in which the magnetic filed depends on the depth of the convective envelope.

The incidence of flare stars of the A-type is 1.16\%. It is much higher than that of F type stars, which goes against the theoretical prediction. The mass range from the A-type star to the F-type star comprise a transition from the radiative envelope to the convective envelope in the exterior layer of a star so that F-type stars can operate a solar-like dynamo, while A-type stars have difficulty to generate and sustain the magnetic field. The discrepancy between the theory and the observation is hard to reconcile, unless we assume that A-type stars have different mechanisms on generating flares.

The FFDs from F-type stars to M-type stars obey a power-law relation with $\alpha \sim 2$. The similarity on the $\alpha$ index of different spectral types implies that they have the same mechanism on flares, provided that the flare is generated by the magnetic field. The FFD of A-type stars has an index $\alpha \sim 1$, which deviates from the common value. The A-type star has a weak magnetic field and barely has a convection envelope.  The energetic flares cannot be associated with the magnetic field by the solar-like dynamo. Both the anomaly of the incidence of flare stars and the deviation of the FFD imply that A-type stars generate flares in a different way.

Since the flare is an indicator for stellar activity, the large sample of flare events and flare stars provide us an opportunity to gauge important issues that connect stellar activity, stellar evolution and stellar structure.
We estimate the flare energy for each flare event and calculate the flare activity ($L_{\rm flare}/L_{\rm bol}$) for each flare star.

The activity--rotation relation in the M-type star is clear, i.e., a saturated regime with a high activity level and a exponential decay regime with a low activity level corresponds to fast rotators and slow rotators respectively. The slope of the unsaturated regime is $\beta \sim $2, which is consistent with previous studies. One interesting fact is that the two regimes become dispersive as the temperature increases.

The age--rotation relation (Gyrochronology) associates the saturated regime and unsaturated regime with convective dynamo and interface dynamo respectively. Combined with the Gyrochronology, we find that the mixing of the two different dynamo stars makes the dispersion. Furthermore, the dependence of the activity of C stars on the color is presented. Based on the dynamos and their correspondence to the activities and the rotations, We propose a scenario on understanding the activity--rotation relation across the H-R diagram. The scenario is illustrated in Section 4.4. We thereby suggest a new expression on the activity--rotation relation in terms of the I vs. C framework of the Gyrochronology, in which the segmentation is on the basis of the dynamo rather than the rotation period.

There are some active giants in the {\it Kepler} mission. However, their activity levels are influenced by stellar parameters. A comparison of stellar parameters between {\it Kepler} and LAMOST shows that the log$g$ of those giants has been underestimated. Most of the active giants are main sequence stars or inactive stars, while few outliers have high activity levels.

We present the rotation distribution of flare stars, which indicates that about 70\% of flare stars rotate faster than 10 days and 95\% of flare stars rotate faster than 30 days. The incidence of flare stars of the G-type is 1.46\% in 4 years, in which 5\% of stars have similar temperature and rotation period with the Sun. Considering the detection limit of the {\it Kepler} mission, we estimate that a superflare with energy $\sim 10^{34}$ occurs on the Sun at least once in 5500 years.

The flare activity has a positive correlation with the size of the starspot. However, about 90\% of stars do not flare, although they have large starspots. On the other hand, some flaring stars do not show apparent variations of lightcurves. It could be attributed to the inclination angle and the modulation of the dynamo. The relation between flares and starspots remain to be addressed by future research.

\begin{figure*}[!htb]
\center
\subfigure{\includegraphics[width=0.9\textwidth]{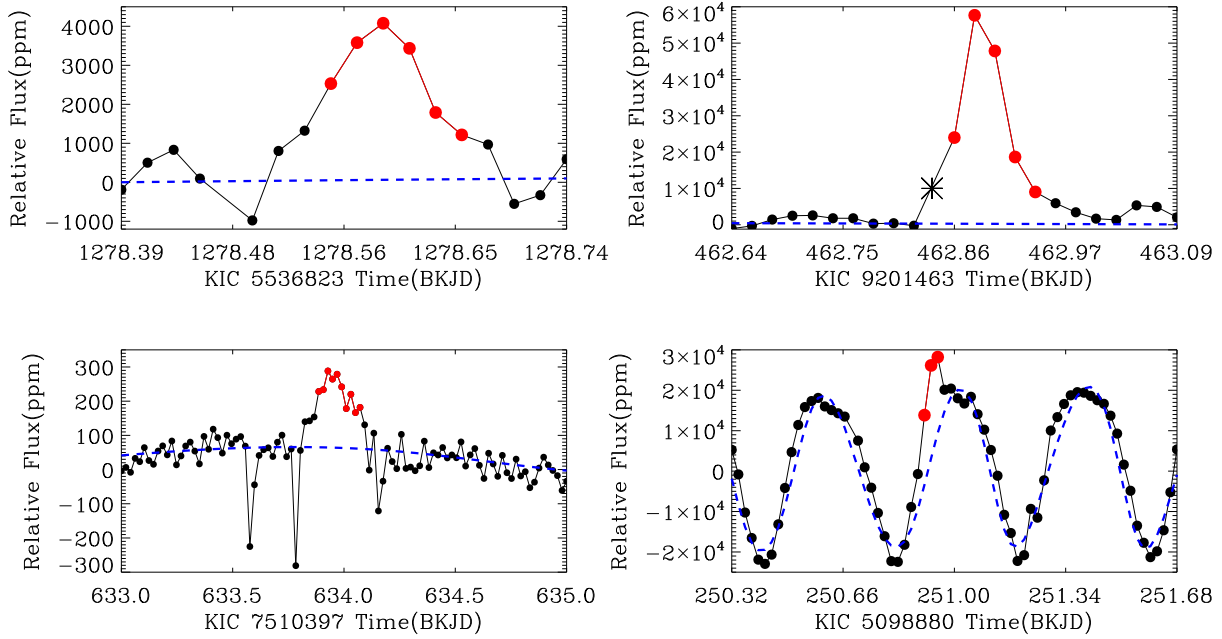}}
\hspace{2 mm}
\caption{The typical deceptive example in the flare detection. The black lines are the raw data. The blue lines are the fitting lines. The red lines represent the flaring segments. The top left panel shows an artifact occurring at the same moment in many stars. The top right panel show an artifact having a break point that marked by an asterisk. The bottom left panel shows an artifact with abnormally low point. The bottom right panel shows an fitting process for a ultra-rapid rotating star.  } \label{figapp}
\end{figure*}

\begin{figure*}[!htb]
\center
\subfigure{\includegraphics[width=0.9\textwidth]{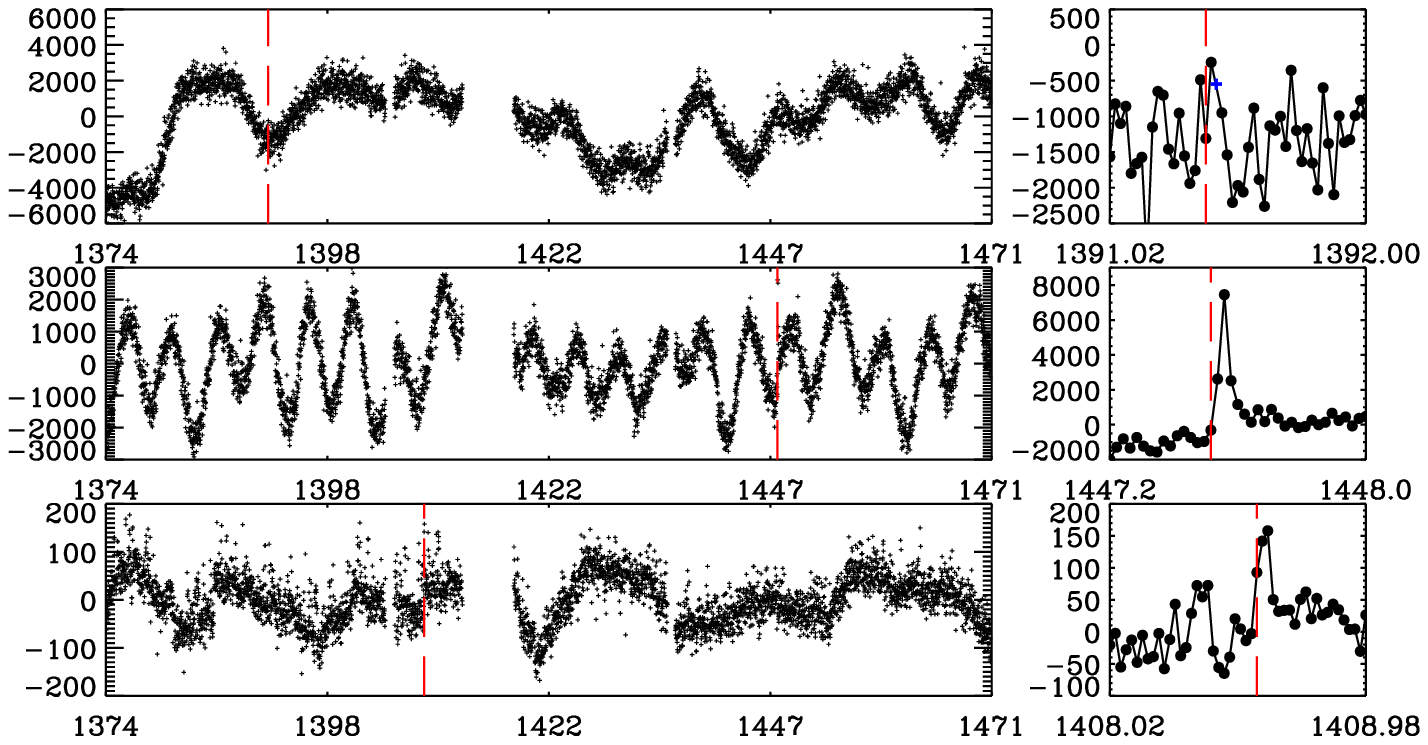}}
\hspace{2 mm}
\caption{Typical disputing cases between this work and \cite{Door2017}. They are in the catalog of \cite{Door2017} but not in this work. The vertical dash line is the begin time of a candidate. The left panels are the overall view of the lightcurve, while the right panels are the details of the flaring segments. The top panel shows a candidate that having a break point (plus symbol). The middle panel shows a star with a unique candidate. The bottom panel shows a candidate in a lightcurve of inferior quality.} \label{figapb}
\end{figure*}

\begin{figure*}[!htb]
\center
\subfigure{\includegraphics[width=0.9\textwidth]{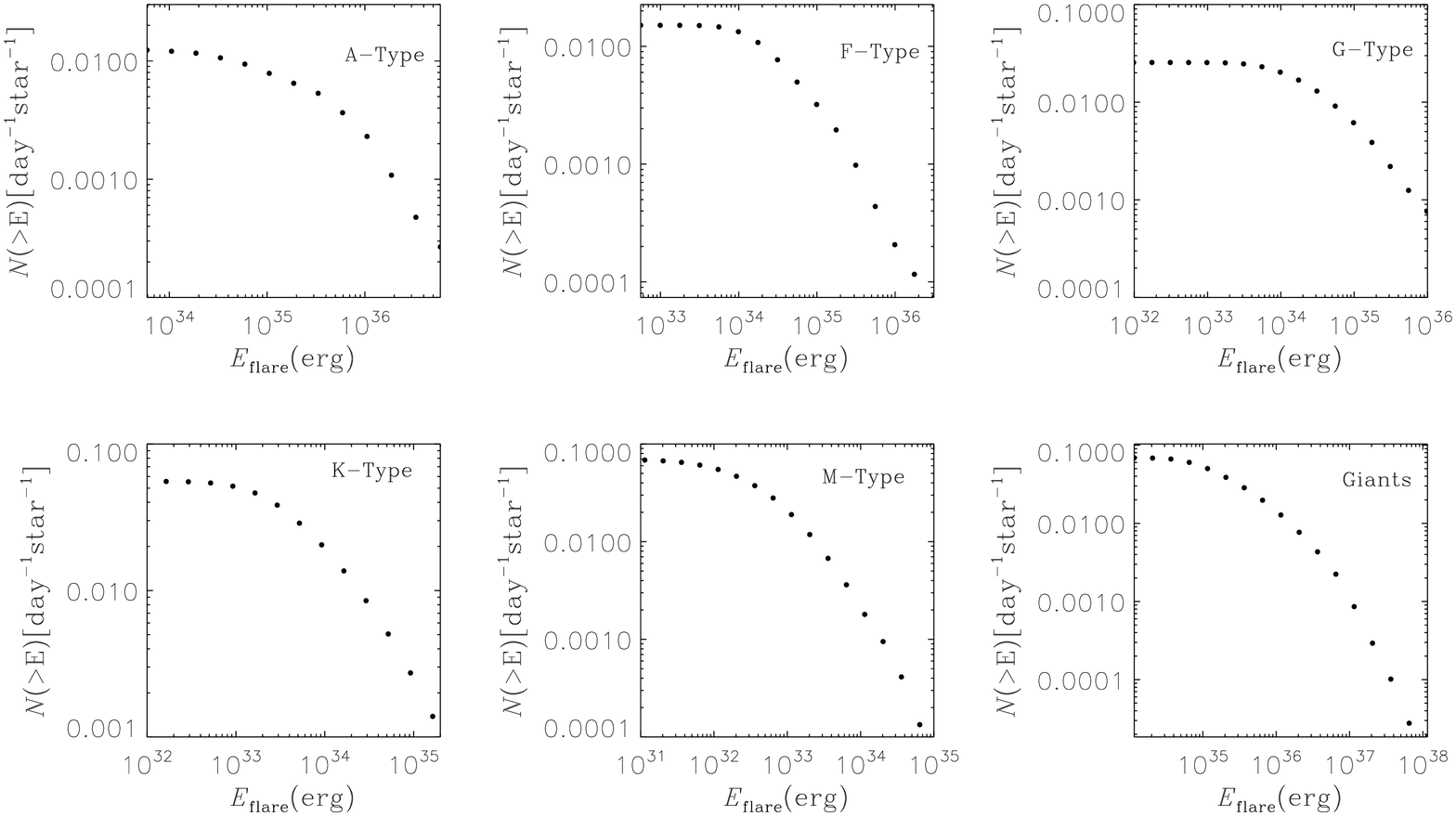}}
\hspace{2 mm}
\caption{The average cumulative FFDs for each type of stars.} \label{figapc}
\end{figure*}

\section*{Appendix A: Deceptive cases in the flare detection}
Besides the pulsating stars, some deceptive cases should be noted in the flare detection. \citet{Yang2018} found that many artifacts occurring at the same moment with similar profiles. All of the candidates in those time segments are excluded. The top left panel of Figure~\ref{figapp} shows an example. In the Kepler data, a set of coadded and stored pixels obtained at a specific time is referred to as a cadence, and the total amount of time over which the data in a cadence is coadded is the cadence period. For LC data, 270 frames are coadded, leading to a total of 1765.5 s = 0.4904 hr \citep{Van Cleve2009}. Sometimes, because of various reasons such as the cooling of CCDs, the count in a cadence period could be a null datum leading to a ``break point" in a continuous observation. The break point in the {\it Kepler} lightcurves could give rise to the re-calibration of the gain of CCDs resulting in false positive signals. All of the candidates that have break points within 3 hours are excluded. The top right panel of Figure~\ref{figapp} shows an example, in which the break point is marked as an asterisk. \citet{Yang2018} found that abnormally low points could cause false positive signals. All of the candidates that have low points within 6 hours are excluded. The bottom left panel shows an example.

For ultra-rapid rotating stars, it is hard to distinguish baselines and outliers. The fitting of the baseline may be tuned manually in the process of the validation and some marginal cases are deemed as flares with arbitrary choices that based on our experience. The bottom right panel shows an example.

\section*{Appendix B: Comparison of marginal cases of flares}

Through the comparison between our catalog and \citet{Door2017}, 53 candidates could be in dispute, although we are inclined to artifacts based on our experience. We present three typical examples here. Figure~\ref{figapb} shows some typical examples that are not in this work. In the top panel, the significance of this candidate is not very prominent, and it has a break point so that it is deemed as an artifact as mentioned in the Appendix A. The middle panel shows a significant candidate, which is the unique one in this star. Its profile is symmetry and unresolved, which does not meet our criteria. The bottom panel shows a candidate in a lightcurve of inferior quality. In the validation, the whole quality of the lightcurve are taken into account. There are many artifacts in the lightcurve, making it unreliable.

\section*{Appendix C: Cumulative FFDs for each type of stars}

We present the average cumulative FFDs for each type of stars in Figure~\ref{figapc}. Values of $y$ axis are calculated as follows:

\begin{center}
\begin{equation}
N_{\rm cum}(>E) = \frac{1}{N_{\rm star}} \sum \frac{N_i(>E)}{t_i}.
\end{equation}
\end{center}

$N_i(>E$) is the number of flares whose energies are larger than $E$ and $t_i$ is the observation time for each flaring star. $N_{\rm star}$ is the number of stars for a given spectral type.

\acknowledgements
\begin{acknowledgements}
We sincerely thank the anonymous referee for the very helpful constructive comments and suggestions, which have significantly improved this article. J.F.L. acknowledges support from the National Science Foundation of China (grant 11425313). The paper includes data collected by the Kepler mission. Funding for the
Kepler mission is provided by the NASA Science Mission Directorate.
All of the data presented in this paper were obtained from the
Mikulsk Archive for Space Telescopes (MAST). STScI is operated by
the Association of Universities for Research in Astronomy, Inc.,
under NASA contract NAS5-26555. Support for MAST for non-HST data is
provided by the NASA Office of Space Science via grant NNX09AF08G
and by other grants and contracts.
\end{acknowledgements}

\end{document}